\documentclass[journal]{IEEEtran}

\newcommand{\figref}[1]{Fig.~\ref{#1}}
\newcommand{\tableref}[1]{Table~\ref{#1}}
\newcommand{\secref}[1]{Section~\ref{#1}}

\ifCLASSINFOpdf
\usepackage[pdftex]{graphicx}
\usepackage{float}
\else
\fi

\usepackage{amsmath}
\usepackage{amssymb}
\usepackage{amsbsy}
\usepackage{amsfonts}%
\usepackage{mathrsfs}
\usepackage{tabu,multirow}
\usepackage{multicol}
\usepackage{float}
\usepackage{subcaption}
\usepackage{array}
\usepackage{verbatim}
\usepackage{color}
\usepackage{colortbl}
\usepackage{flushend}

\begin{document}
		\title{Deep Optimization of Parametric IIR Filters for Audio Equalization}
	
	\author{Giovanni~Pepe,~Leonardo~Gabrielli,~Stefano~Squartini,~Carlo~Tripodi,~and~Nicol\`{o}~Strozzi%
		\thanks{G. Pepe is with the Department of Information Engineering, Universit\`a Politecnica delle Marche, Via Brecce Bianche, 60131, Ancona (AN), Italy and with ASK Industries S.p.A., Viale Ramazzini, 42124, Reggio Emilia (RE), Italy, e-mail: g.pepe@pm.univpm.it.}
		\thanks{L. Gabrielli and S. Squaritni are with the Department of Information Engineering, Universit\`a Politecnica delle Marche, Via Brecce Bianche, 60131, Ancona (AN), Italy.}%
		\thanks{C. Tripodi and N. Strozzi are with ASK Industries S.p.A., Viale Ramazzini, 42124, Reggio Emilia (RE), Italy.}
	}
	\markboth{IEEE/ACM TRANSACTIONS ON AUDIO, SPEECH, AND LANGUAGE PROCESSING}%
	{Shell \MakeLowercase{\textit{et al.}}: Bare Demo of IEEEtran.cls for IEEE Journals}
	
	\maketitle
	
	\begin{abstract}
This paper describes a novel Deep Learning method for the design of IIR parametric filters for automatic audio equalization. A simple and effective neural architecture, named BiasNet, is proposed to determine the IIR equalizer parameters. An output denormalization technique is used to obtain accurate tuning of the IIR filters center frequency, quality factor and gain. All layers involved in the proposed method are shown to be differentiable, allowing backpropagation to optimize the network weights and achieve, after a number of training iterations, the optimal output. The parameters are optimized with respect to a loss function based on a spectral distance between the measured and desired magnitude response, and a regularization term used to achieve a spatialization of the acoustc scene. Two scenarios with different characteristics were considered for the experimental evaluation: a room and a car cabin. The performance of the proposed method improves over the baseline techniques and achieves an almost flat band. Moreover IIR filters provide a consistently lower computational cost during runtime with respect to FIR filters.
	\end{abstract}
	
	
	\IEEEpeerreviewmaketitle

\section{Introduction}
\IEEEPARstart{P}{arametric} equalizers are among the most used equalizers in audio and acoustics, thanks to the few parameters to control: center frequency, gain and quality factor \cite{vesa2016equalization}, which boost or cut the spectrum. Such equalizers are usually composed of a cascade of Second Order Sections (SOS's) Infinite Impulse Response (IIR) filters, designed according to some constraints, such as a unitary gain at Direct Current (DC) and Nyquist frequency \cite{reiss2011design}.

Audio equalization improves sound quality in listening environments, such as rooms, halls, car's cabins and so on \cite{Karjalainen2005,cecchi2018room}.
The in-car sound scene is changing in the last years \cite{environments5040044}, improving acoustic insulation and absorption and, thanks to hybrid and electric engines, leading to an interior sound quieter than the combustion vehicles \cite{HUANG201998}. Therefore audio equalization for automotive applications can improve the user experience inside the car's cabin. Audio enhancement in the automotive industry is generally done using Digital Signal Processing (DSP) systems \cite{cecchi2009automotive}, which get cheaper and more performing in the years.

Machine Learning techniques, however, offer an alternative approach with respect to standard approaches and they are progressively adopted in  audio equalization. Early works employing neural networks to invert a room response, can be found e.g. in \cite{chang1994inverse}, while recent end-to-end approaches have been proposed in \cite{martinez2018equalization}. Completely neural-based algorithms, however, require a dedicated hardware to run in real-time. Currently, embedded Tensor Processing Units (TPU) are getting commercially available as embedded devices, paving the way for the implementation of neural audio processing techniques in many application scenarios. 

On the other hand, DSP chips are by far more convenient and energy-efficient for usage in audio equalization. For this reason, Machine Learning techniques may be used for the offline design of linear filters and then implemented in real time on a DSP chip. In literature, several IIR filter design techniques are present \cite{AGRAWAL2021107669}, e.g., using linear programming \cite{krusienski2003pso}, non-linear optimization methods \cite{saha2014gsa,dodds2020a} and evolutionary algorithms \cite{jiang2010}. These are gradient-free algorithms \cite{Sloss2020}, and among them, Particle Swarm Optimization (PSO) and Gravitational Search Algorithm (GSA) \cite{nongpiur2013} are the most widely used.

Recent works started employing neural networks for the design of IIR filters and parametric equalizers, mainly following the idea of Differentiable Digital Signal Processing (DDSP) layers \cite{Engel2020DDSP}. These layers can backpropagate the error and are, thus, amenable for integration in a Machine Learning framework. In \cite{kuznetsovdifferentiable,Bhattacharya2020OPTIMIZATIONOC} the concept of adaptive filtering using state-of-the-art backpropagation algorithms have been presented, removing the convex optimization constraints imposed by different algorithms such as the least-squares approach which is widely used in classic adaptive filter theory. Another work which adopts the DDSP approach is introduced in \cite{Nercessian2020Neural}, where a neural network is trained to estimate the equalizing filter coefficients from a desired frequency response. Earlier works dealing with backpropagation for the design of IIR and Finite Impulse Response (FIR) filters are described in \cite{back1991synapses,Allakhverdiyeva2016nnfilter}.
Finally, the works in \cite{valimaki2019,ramo2019neural} deal with the correction of octave and third-octave graphic equalizers gain, in order to reduce the error introduced by the interaction of neighboring filters, and achieve an accurate prediction of the equalizer gains with a feedforward neural network.

\subsection{Scope of the work} \label{sec:scope}

Current filter design approaches based on the backpropagation of the error address the system identification problem \cite{kuznetsovdifferentiable,Bhattacharya2020OPTIMIZATIONOC} or frequency response matching \cite{Nercessian2020Neural}. 
Our work presents two main differences with respect to the above techniques: the goal and the method. 

Our goal is to design filters for room equalization, therefore the neural network does not have a reference frequency response to directly match. Instead it must estimate a proper approximation of the room inverse and identify the related filter coefficients. This task is challenging as the network must first approximate the room inverse response and then find suitable filter coefficients. Furthermore, we aim to obtain a room equalization not only in the Single-Input Single-Output (SISO) case, but also for any number of sound sources and listening points (Multiple-Input Multiple-Output or MIMO), which arguably makes the task more challenging. We assume that the test environment is static, enabling an offline filter design. Factors that may change the Room Impulse Responses (RIRs) over time (such as changes in the environment geometry, listener positions, temperature change, etc.) may impair the performance, and must be addressed using other techniques, such as adaptive filtering, not covered in this paper. Our technique is not suitable for online adaptive equalization due to the computing time required for the filter design, but it is very useful in the pre-tuning phase, which always plays an important role in equalization of real acoustic environments.

As for the method, we employ neural networks with an unconventional  approach, that we will refer to as Deep Optimization (DO), owing the name from the use of the neural network as an non-convex optimization algorithm. This approach stems from our previous works in equalizing filter design \cite{PEPE2020107204,pepe2020designing}. However, here it is investigated in more detail and declined in a simpler and more efficient way. 

In our previous works, we first tackled room equalization filter design using evolutionary algorithms in \cite{PEPE2020107204}, where FIR filters were designed in an automotive MIMO scenario. In \cite{pepe2020designing}, we largely improved the audio equalization performance by exploiting neural networks and outperforming the traditional Frequency Deconvolution (FD) method \cite{kirkeby1998freq} by several orders of magnitude. This new filter design method opened several research questions related to network design that we are trying to address here. Furthermore, the design of FIR filters has some downsides: they have a high computational cost, and cannot be hand-tuned, which is an often desired feature, since listeners may require a personalized audio experience. To overcome this, IIR equalizers can be used. These provide a lower computational cost and the possibility of hand tuning after the automatic filter design is completed. 
Automatic IIR parametric equalizers were first discussed in \cite{ramos2006filter}, where the Direct Search Method (DSM), a recursive technique, is used to optimize the parameters. The Rosenbrock method was presented in \cite{behrends2011automatic}, where it was used to equalize an LCD TV speaker. We applied both method to the room equalization problem in \cite{pepegravitational}, and compared to GSA. The GSA resulted in superior performance, but still far from the performance obtained by the deep neural design of FIR filters proposed in \cite{pepe2020designing}. For this reason, deep neural techniques are worth investigating for the design of an IIR equalizer for room equalization that can match the performance previously obtained with FIR filters.

The present work is organized as follows: in Section \ref{sec:problem} we introduce the MIMO room equalization problem in formal terms and IIR parametric filters transfer functions and coefficients design. In Section \ref{sec:proposed} we propose a novel neural network architecture for the design of IIR equalizers for room equalization, along with a novel denormalization technique and loss function. In Section \ref{sec:baseline} we introduce two alternative neural architectures and two baseline techniques for comparison purposes. In Sections \ref{sec:exp} and \ref{sec:results} report, respectively, two experimental setups and their results are introduced. Finally, Section \ref{sec:conclusion} concludes the paper.
 
\section{Problem Statement}
\label{sec:problem}

As previously stated, our main goal is to achieve room equalization with a parametric IIR equalizer, which is composed of a cascade of $N$ second order digital IIR filters of the biquad type, i.e. SOS's. Parametric filters can be of the \textit{peaking} or \textit{shelving} type \cite{zolzer2011dafx}. The former introduces a gain $V_{0}$ at a specific frequency $f_c$. The gain around $f_c$ goes to 1 with a slope depending on the quality factor $Q$. The shelving filter, instead, introduces a flat gain starting from or up to a specific frequency. In this work we restrict our attention to peaking filters to avoid unnecessary equalization at the extremes of the hearing range. Their coefficients are computed according to a closed form expression with its three free parameters. In our work, additionally, we include another free parameter, a \textit{channel gain} $V_{s}$, one for each audio channel, i.e. sound source. Each equalizer can be expressed as a product of rational functions in the Z-domain (the individual SOS's transfer functions), i.e.:
\begin{equation}
G_s(z)=V_s\prod_{\kappa=1}^{N}\frac{b_{0,s,\kappa}+b_{1,s,\kappa}z^{-1}+b_{2,s,\kappa}z^{-2}}{a_{0,s,\kappa}+a_{1,s,\kappa}z^{-1}+a_{2,s,\kappa}z^{-2}}	
\end{equation}
where $G_s(z)$ is the transfer function of a parametric IIR filter of the $s$-th speaker, $\kappa$ is the index of $\kappa$-th  SOS, $N$ is the number of SOS's and $V_s$ is the channel gain in linear scale.

\begin{figure}[h!tb]
	\centering
	\includegraphics[width=\linewidth,,trim={1.0cm 1cm 1cm 21cm},clip]{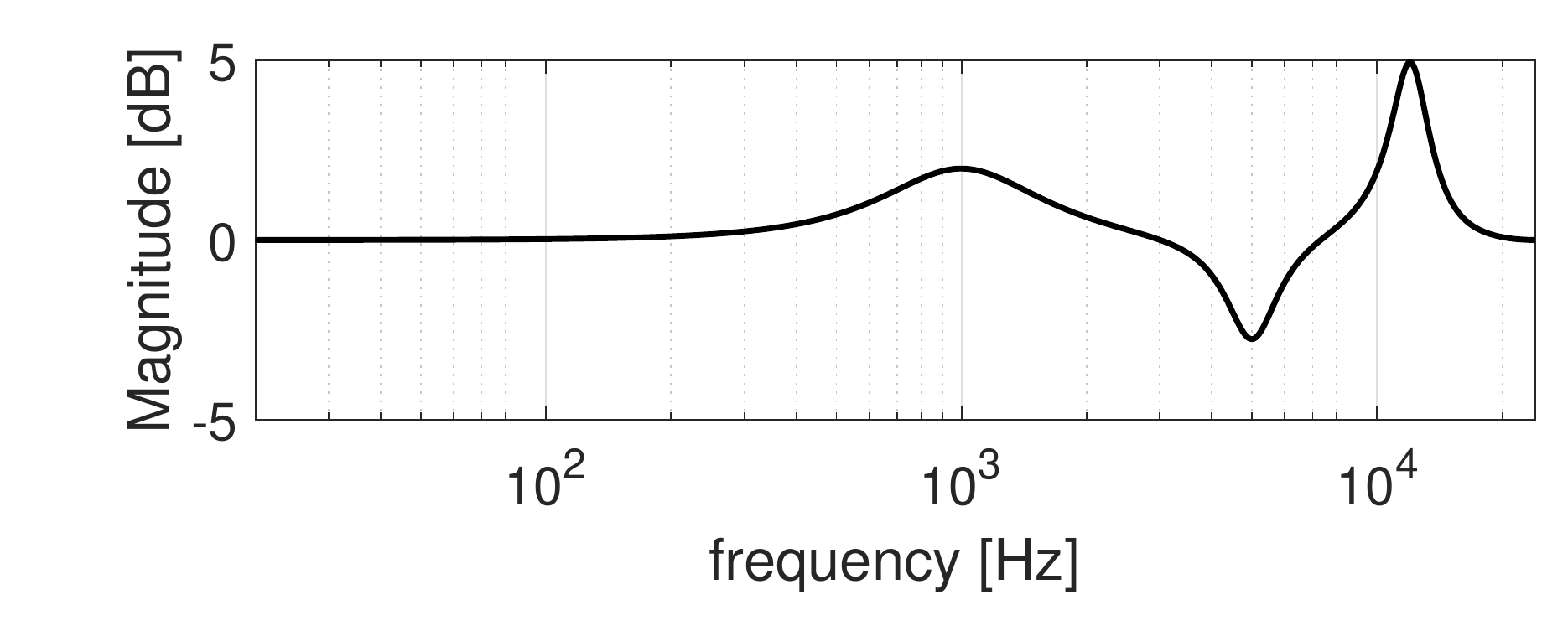}
	\caption{Example of a parametric IIR filter composed of three SOS's. The center frequencies are: 1\,kHz, 5\,kHz and 12\,kHz.}
	\label{fig:example_par}
\end{figure}

The coefficients are calculated according to the type of filter (boost or cut) as described in \tableref{table:coeff_iir}.  A parametric IIR filter composed of three SOS’s is shown in \figref{fig:example_par} as an example.

Overall, the MIMO room equalization system is composed of $M$ listening points and $\mathcal{S}$ sound sources. Each sound source has an associated IIR equalizer, composed of $N$ SOS's. Clearly, the SISO case can be regarded as a special case with $\mathcal{S} =1, \mathcal{M}=1$. 
The measured signal at the m-th listening position is
\begin{equation} \label{eq:signal_listening}
y_m(n)=\sum_{s=1}^{\mathcal{S}} x(n) * g_s(n) * h_{s,m}(n),
\end{equation}
where $g_s(n)$ is the impulse response of the s-th equalizer, $h_{s,m}(n)$ is the impulse response between the m-th point and the s-th source and $x(n)$ is an audio input signal. 

The objective of an automatic room equalizer using parametric IIR filters is to determine the equalizer parameters $f_c, Q, V_{0}, V_{s}$ that achieve a flat (or any other desired) frequency response at the listening points. This can be seen as an optimization problem, as discussed in the following section.

\begin{table*}[h!tb]
	\centering
	\renewcommand{\arraystretch}{1.3}
	\caption{Coefficients calculated according to the boost or cut filter \cite{zolzer2011dafx}.}
	\begin{tabular}{c|c||c}
		\hline
		IIR Coefficients&Boost ($V_{0_{s,\kappa}}\geq 1$)& Cut ($0 < V_{0_{s,\kappa}}< 1$)\\\hline\hline
		$b_{s,\kappa,0}$&$\frac{1+\frac{{V_{0,s,\kappa}}}{Q_{s,\kappa}}\cdot tan(\pi\cdot f_{c_{s,\kappa}}/f_s)+tan(\pi\cdot f_{c_{s,\kappa}}/f_s)^2}{1+1/Q_{s,\kappa}\cdot tan(\pi\cdot f_{c_{s,\kappa}}/f_s)+tan(\pi\cdot f_{c_{s,\kappa}}/f_s)^2}$&$\frac{1+\frac{1}{Q_{s,\kappa}}\cdot tan(\pi\cdot f_{c_{s,\kappa}}/f_s)+tan(\pi\cdot f_{c_{s,\kappa}}/f_s)^2}{1+{V_{0,s,\kappa}}/Q_{s,\kappa}\cdot tan(\pi\cdot f_{c_{s,\kappa}}/f_s)+tan(\pi\cdot f_{c_{s,\kappa}}/f_s)^2}$\\\hline\hline
		$b_{s,\kappa,1}$&$\frac{2\cdot tan(\pi\cdot f_{c_{s,\kappa}}/f_s)^2-1}{1+1/Q_{s,\kappa}\cdot tan(\pi\cdot f_{c_{s,\kappa}}/f_s)+tan(\pi\cdot f_{c_{s,\kappa}}/f_s)^2}$&$\frac{2\cdot tan(\pi\cdot f_{c_{s,\kappa}}/f_s)^2-1}{1+{V_{0,s,\kappa}}/Q_{s,\kappa}\cdot tan(\pi\cdot f_{c_{s,\kappa}}/f_s)+tan(\pi\cdot f_{c_{s,\kappa}}/f_s)^2}$\\\hline\hline
		$b_{s,\kappa,2}$&$\frac{1-\frac{{V_{0,s,\kappa}}}{Q_{s,\kappa}}\cdot tan(\pi\cdot f_{c_{s,\kappa}}/f_s)+tan(\pi\cdot f_{c_{s,\kappa}}/f_s)^2}{1+1/Q_{s,\kappa}\cdot tan(\pi\cdot f_{c_{s,\kappa}}/f_s)+tan(\pi\cdot f_{c_{s,\kappa}}/f_s)^2}$&$\frac{2\cdot tan(\pi\cdot f_{c_{s,\kappa}}/f_s)^2-1}{1+{V_{0,s,\kappa}}/Q_{s,\kappa}\cdot tan(\pi\cdot f_{c_{s,\kappa}}/f_s)+tan(\pi\cdot f_{c_{s,\kappa}}/f_s)^2}$\\\hline\hline
		$a_{s,\kappa,0}$&1&1\\\hline\hline
		$a_{s,\kappa,1}$&$\frac{2\cdot tan(\pi\cdot f_{c_{s,\kappa}}/f_s)^2-1}{1+1/Q_{s,\kappa}\cdot tan(\pi\cdot f_{c_{s,\kappa}}/f_s)+tan(\pi\cdot f_{c_{s,\kappa}}/f_s)^2}$&$\frac{2\cdot tan(\pi\cdot f_{c_{s,\kappa}}/f_s)^2-1}{1+{V_{0,s,\kappa}}/Q_{s,\kappa}\cdot tan(\pi\cdot f_{c_{s,\kappa}}/f_s)+tan(\pi\cdot f_{c_{s,\kappa}}/f_s)^2}$\\\hline\hline
		$a_{s,\kappa,2}$&$\frac{1-\frac{1}{Q_{s,\kappa}}\cdot tan(\pi\cdot f_{c_{s,\kappa}}/f_s)+tan(\pi\cdot f_{c_{s,\kappa}}/f_s)^2}{1+1/Q_{s,\kappa}\cdot tan(\pi\cdot f_{c_{s,\kappa}}/f_s)+tan(\pi\cdot f_{c_{s,\kappa}}/f_s)^2}$&$\frac{1-\frac{{V_{0,s,\kappa}}}{Q_{s,\kappa}}\cdot tan(\pi\cdot f_{c_{s,\kappa}}/f_s)+tan(\pi\cdot f_{c_{s,\kappa}}/f_s)^2}{1+{V_{0,s,\kappa}}/Q_{s,\kappa}\cdot tan(\pi\cdot f_{c_{s,\kappa}}/f_s)+tan(\pi\cdot f_{c_{s,\kappa}}/f_s)^2}
		$\\\hline
	\end{tabular}
	\label{table:coeff_iir}
\end{table*}

\section{Proposed Method} \label{sec:proposed}

\begin{figure}
	\centering
	\includegraphics[width=1\textwidth,trim={0cm 0cm 3cm 0cm},clip]{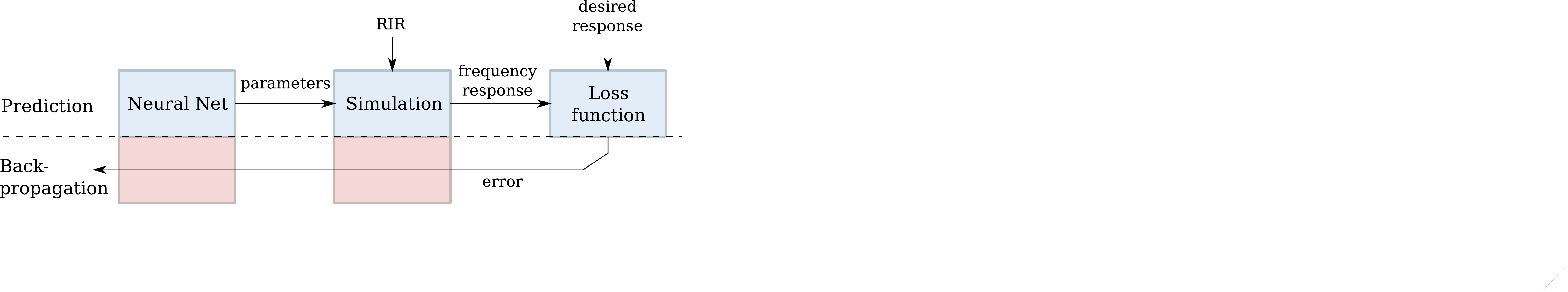}
	\caption{Overview of Deep Optimization for room equalization. The algorithm consists of a forward pass (top) to predict equalizer parameters and the backpropagation of the error (bottom) to optimize the performance iteratively. The red boxes perform the backward flow of the partial derivatives of the operations performed by the blue boxes.}
	\label{fig:general}
\end{figure}

As described in \secref{sec:scope}, in this work we propose a novel and simpler Deep Optimization neural network architecture for the design of room equalization IIR parametric filters.
As demonstrated in \cite{lopez2018easing}, deep neural networks can solve non-convex problems through the minimization of a cost function. The design of equalizing filters can be formalized as such. This involves employing the neural network in a different way with respect to common classification and regression problems. The proposed approach is depicted in \figref{fig:general}: the neural network initially computes filter coefficients based on random initial values of the weights. The coefficients are used, in conjunction with the room impulse responses to predict the frequency response obtained at one or more listening points. This is used to calculate the error with respect to the desired frequency response according to a given loss function. The error is then backpropagated to optimize the network weights in order to iteratively produce a more accurate frequency response with respect to the desired one.


In \cite{pepe2020designing}, we employed a convolutional neural network for FIR filter design for audio equalization. This, however, requires substantial computing power if the number of neurons in the convolutional layers is large. Their number is constrained by the size of the input. In absence of prior works, the neural network employed in \cite{pepe2020designing} was fed with a three-dimensional input tensor containing the RIRs. This however, was shown to be not necessary for the network: any input of the same dimensions and size filled with non-zero values would achieve similar results. More specifically, a tensor of 1s would lead to slightly inferior results, while changing the input with random values at each iteration would result in a slow convergence of the network. Our experiments, discussed as part of this paper, allow concluding that the network does not require any informative content at its input. The only information that is required to perform an effective equalization is the knowledge of the room acoustic response in terms of RIR, used in the simulation stage.
Building on these observations, in this work we propose BiasNet, an efficient neural network for optimization which avoids the need of convolutional input layers.

As described previously, the architecture must be also designed to cope with IIR filters. In the FIR design task, the neural network directly computes the filters coefficients. This strategy is not suitable for IIR filters, since recursive filter poles are subject to stability and quantization issues \cite{pepegravitational}. We choose to estimate the parameters of a parametric IIR equalizer, then compute the IIR coefficients according to the closed form equations of Table \ref{table:coeff_iir}. The output layer of the network is, thus, composed of $(3 \times N)\times \mathcal{S} + \mathcal{S}$ neurons. As an additional novelty, the parameters are then denormalized by using linear interpolation in their defined range. However, to avoid overlapping filters and reduce the center frequency prediction error of each SOS, we propose a constrained denormalization of the center frequency parameter $p_{f_c,\kappa} \in [-1,1]$, which is related to the center frequency of the $\kappa$-th SOS as follows:
\begin{equation}\label{eq:fcdenorm}
	f_c=\frac{f_{c,max}-f_{c,min}}{2}\cdot p_{f_c,\kappa} +\frac{f_{c,max}+f_{c,min}}{2}
\end{equation}
where $f_{c,max}$ and $f_{c,min}$ are the maximum and minimum allowed values for the center frequency. These values can be devised, e.g., according to third-octave bands, or any other subdivision of the audio range. This subdivision constraints the number of SOS's to the number of frequency bands, avoiding the overlap of the filters operative bandwidth which, in turn, may result in excessive gains for some bands. Furthermore, mapping the range $[-1, 1]$ to a narrow portion of the spectrum reduces the prediction error of the $f_c$. The other parameters are denormalized mapping $[-1, 1]$ to their full range which can be defined according to the application. Gains are designed by the network on a dB scale and are then converted into linear values when computing the IIR biquad equations of Table \ref{table:coeff_iir}. From now on, dB scale gains will be denoted as $V_{0,dB}, V_{s,dB}$ to avoid confusion with their linear counterparts $V_0, V_s$.

After computing the IIR filters, these are used to simulate the frequency response at the desired listening positions employing the related impulse responses. With respect to \cite{Bhattacharya2020OPTIMIZATIONOC}, instead of using convolution in time we compute the product in the frequency domain \cite{Oppenheim} due to its reduced computational cost. This optimization is important to reduce the overall computing time, as the optimization process involves a large number of iterations.

The frequency response of the numerator and the denominator of a SOS is given by:
\begin{equation}
	B_{s,\kappa}(k)=\mathcal{F}[b_{s,\kappa}(z)]=b_{0_{s,\kappa}}+b_{1_{s,\kappa}}e^{-\frac{j 2\pi k}{N}}+b_{2_{s,\kappa}}e^{-\frac{j 2\pi 2 k}{N}}
\end{equation}
\begin{equation}
	A_{s,\kappa}(k)=\mathcal{F}[a_{s,\kappa}(z)]=a_{0_{s,\kappa}}+a_{1_{s,\kappa}}e^{-\frac{j 2\pi k}{N}}+a_{2_{s,\kappa}}e^{-\frac{j 2\pi 2 k}{N}}
\end{equation}

The equalized frequency response $\tilde{H}_{s,m}(k)$ between the $s$-th speaker and the $m$-th microphone is given by:
\begin{equation}\label{eq:eq_frequency}
	\tilde{H}_{s,m}(k)=H_{s,m}(k)\cdot V_{s}\prod_{\kappa=1}^{K}\frac{B_{s,\kappa}(k)}{A_{s,\kappa}(k)}
\end{equation}
where $H_{s,m}(k)$ is the frequency response achieved from the RIR by the discrete Fourier transform.

All the operations described above (denormalization, closed-form IIR coefficients computing, FFT filtering, etc.) must be differentiable in order to be employed in a neural network setting and to allow error backpropagation \cite{Engel2020DDSP}. This will be discussed in \secref{sec:back}.

\subsection{BiasNet} \label{sec:ni}

\begin{figure}[h!tb]
	\centering
	\includegraphics[width=\linewidth]{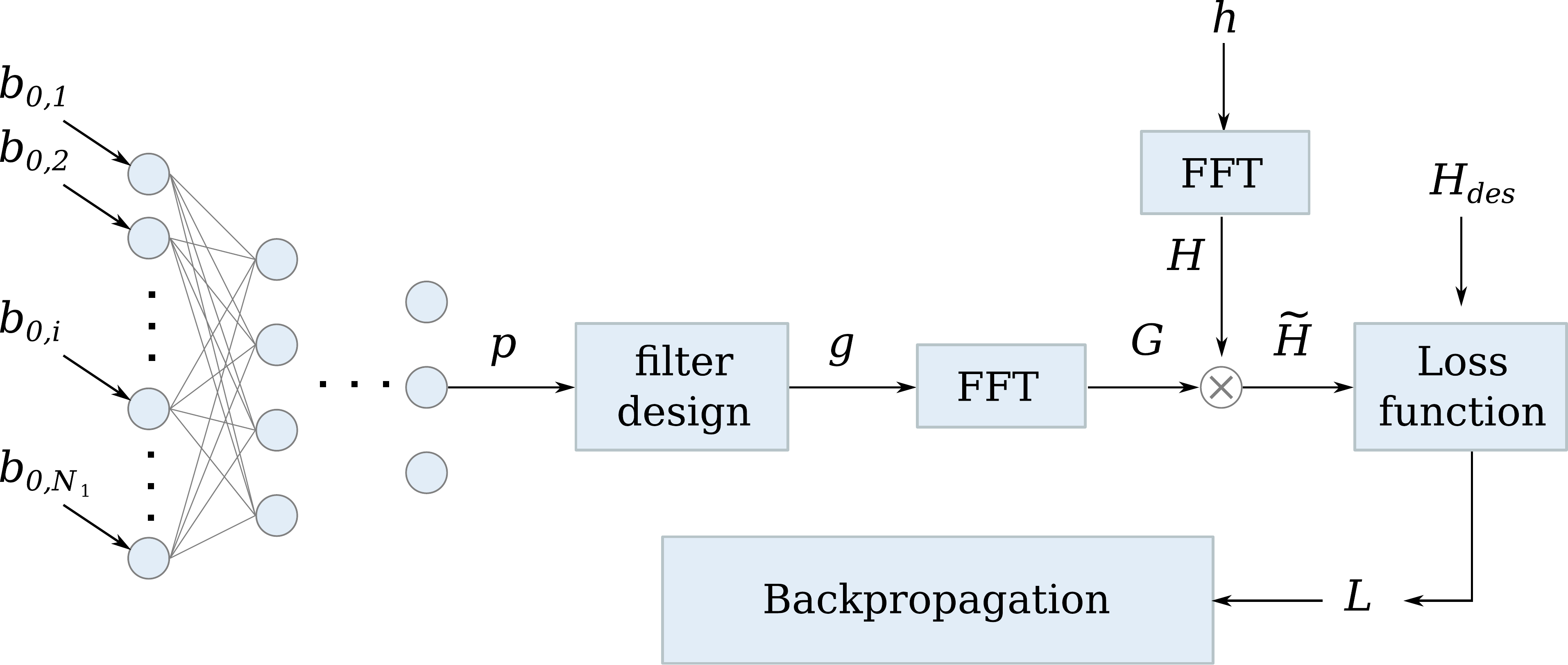}
	\caption{BiasNet overview: the network provides parameters $p$ that are denormalized and used of the filter design. The filters are transformed in the frequency domain, then frequency convolution is performed to compute the equalized response, that is compared with the desired response to compute the loss value. This is used in backpropagation to learn the optimal network weights and input bias terms.}
	\label{fig:bias_scheme}
\end{figure}

Within the DO scenario, a neural network can be seen as a unidirectional graph of nonlinear computations. If the input is fixed and only the network weights can change, then convergence is only determined by the update of the network weights, and is sufficiently fast as we shall see later. As shown in \cite{pepe2020designing}, if the activation functions satisfy the relation $f(0) = 0$, an input tensor with 0s will not produce any output, as the result of the computations will be, similarly, 0s. Therefore, some energy must be injected into the system. We do so by taking advantage of the bias terms of the input layer. These are learnable parameters, therefore, they will be updated during the optimization procedure. The vector of input neurons' bias terms $\mathbf{b_0}$ can be seen, i.e., as a learnable input vector. 
The proposed architecture is called BiasNet, and consists of a feedforward network with no input but a learnable bias term, as shown in \figref{fig:bias_scheme}.

The advantage of using this architecture, with respect to the convolutional neural networks used previously, lies in: \textit{(a)} the learnable input, which provides an optimal solution and avoids tweaking the input size and content; and \textit{(b)} the low number of network parameters to be learned, which influences the convergence speed, as we shall see later.

\subsection{Loss function}

The cost function has a salient role in the optimization process since it computes the error and thus drives the network weights update procedure. The loss used in this work is a combination of a spectral loss $L_1$, and a multichannel energy regularization term $L_2$:
\begin{equation}\label{eq:loss}
L= \gamma_1 \cdot L_1+\gamma_2\cdot L_2
\end{equation}
with $\gamma_1=1$ and $\gamma_2=log_2(\mathcal{S)}+log_2(\mathcal{M})$ being weights.

The loss function $L_1$ is given by the Euclidean distance between the simulated magnitude response and the desired magnitude response $|H_{des}(k)|$:
\begin{equation}\label{eq:l1}
L_1=\sum_{m=1}^{\mathcal{M}} \sqrt{\sum_{k} (|\tilde{H}_{m}(k)|-|H_{des}(k)|)^2}
\end{equation}
where: $\tilde{H}_m(k)$ is the Discrete Fourier Transform of the equalized impulse response at the m-th microphone $\tilde{h}_m(n)$.

The regularization term $L_2$ is required when $\mathcal{S} >1$ to keep the original energy balance between a reference speaker and each other speaker, avoiding unwanted change in spatial perception. The term is defined as:
\begin{equation}\label{eq:l2}
L_2=\sum_{m=1}^{\mathcal{M}}\sqrt{\sum_{s=1}^{\mathcal{S}}(\hat{r}_{s,m}-r_{s,m})^2}
\end{equation}
where $r_{s,m}$ and $\tilde{r}_{s,m}$ are the ratio between the energy of a reference speaker and the $s$-th speaker before and after equalization, respectively.

\subsection{Differentiating the DSP operations} \label{sec:back}

Backpropagation is performed using the partial derivative of the loss function with respect to the control parameters $\partial L/\partial f_c$, $\partial L/\partial Q$, $\partial L/\partial V_{0,dB}$ and $\partial L/\partial V_{s,dB}$. The partial derivatives are calculated as the product of cascaded local ones. While these are well known for all the typical neural network layers used in this work, for the sake of completeness, we need to develop the mathematical expressions of the partial derivatives of all the DSP operations that allow to simulate the equalized RIRs, i.e.: those in Table \ref{table:coeff_iir}, Eq. \ref{eq:eq_frequency} and so on.

In order to obtain the partial derivative of the loss function with respect to the control parameters, we first need to calculate the partial derivative of the loss function with respect to a generic filtered impulse response:

\begin{equation}
	\frac{\partial L}{\partial \tilde{h}_{s,m}}=\frac{\partial L}{\partial L_1}\cdot\frac{\partial L_1}{\partial \tilde{h}_{s,m}}+\frac{\partial L}{\partial L_2}\cdot\frac{\partial L_2}{\partial \tilde{h}_{s,m}}
\end{equation}
where, from Eq. \eqref{eq:loss}, $\frac{\partial L}{\partial L_1}=\gamma_1$ and $\frac{\partial L}{\partial L_2}=\gamma_2$.

The partial derivative $\partial L_1/\partial \tilde{h}_{s,m}(n)$ is given by the product of:
\begin{equation}
	\frac{\partial L_1}{\partial \tilde{h}_{s,m}(n)}=\frac{\partial L_1}{\partial \tilde{h}_m(n)}\cdot \frac{\partial \tilde{h}_m(n)}{\partial \tilde{h}_{s,m}(n)}.
\end{equation}
Knowing that at one listening point $\tilde{h}_m(n)=\sum_{s=1}^{\mathcal{S}} \tilde{h}_{s,m}(n)$, we can state that $\frac{\partial \tilde{h}_m(n)}{\partial \tilde{h}_{s,m}(n)}=1$, thus $	\frac{\partial L_1}{\partial \tilde{h}_{s,m}(n)}=\frac{\partial L_1}{\partial \tilde{h}_m(n)}$.
Since the filtered room response is computed in the frequency domain we expand Eq. \ref{eq:l1} and use Wirtinger calculus \cite{caracalla2017gradient} to get:

\begin{equation}
	\resizebox{\linewidth}{!}{$
		\begin{split}
			\frac{\partial L_1}{\partial \tilde{h}_m(n)}=&\sum_{k=0}^{N-1}\Bigg[\frac{|\tilde{H}_m(k)|-|H_{d,m}(k)|}{\sqrt{\sum_{k} (|\tilde{H}_{m}(k)|-|H_{d,m}(k)|)^2}} \cdot \frac{Re[\tilde{H}_{m}(k)]}{|\tilde{H}_{m}(k)|} \cdot cos(\frac{2\pi}{N}k n)\Bigg]\\ &-\sum_{k=0}^{N-1}\Bigg[\frac{|\tilde{H}_m(k)|-|H_{d,m}(k)|}{\sqrt{\sum_{k} (|\tilde{H}_{m}(k)|-|H_{d,m}(k)|)^2}} \cdot \frac{Im[\tilde{H}_{m}(k)]}{|\tilde{H}_{m}(k)|} \cdot sin(\frac{2\pi}{N}k n)\Bigg]
		\end{split}
		$}
\end{equation}

To calculate the local derivative of $\tilde{h}_{s,m}(n)$ with respect to a generic control parameter $p$, we first use the Wirtinger calculus \cite{caracalla2017gradient} to determine the local derivative:
\begin{equation}
	\frac{\partial \tilde{h}_{s,m}(n)}{\partial \tilde{H}_{s,m}(k)}=\sum_{n=0}^{N-1} cos\Big(2\pi \frac{k n}{N}\Big)+sin\Big(2\pi \frac{k n}{N}\Big)	
\end{equation}

The partial derivative of the magnitude response with respect to the channel gain $\frac{\partial \tilde{H}_{s,m}(k)}{\partial V_{s,dB}}$ is:
\begin{equation}
	\resizebox{\linewidth}{!}{$
		\frac{\partial \tilde{H}_{s,m}(k)}{\partial 
			V_{s,dB}}=H_{s,m}(k)\cdot \frac{log(10)\cdot 10^{V_{s,dB}/20}}{20}\prod_{j=1}^{K}\frac{B_{s,j}(k)}{A_{s,j}(k)}
		$}
\end{equation}

For the other parameters, we first calculate the partial derivative of $\frac{\partial \tilde{H}_{s,m}(k)}{\partial B_{s,\kappa}(k)}$ and $\frac{\partial \tilde{H}_{s,m}(k)}{\partial A_{s,\kappa}(k)}$:
\begin{equation}
	\resizebox{\linewidth}{!}{$
		\frac{\partial \tilde{H}_{s,m}(k)}{\partial B_{s,\kappa}(k)}=H_{s,m}(k)\cdot V_{s}\cdot\frac{1}{A_{s,\kappa}(k)}\prod_{j=1,j\neq \kappa}^{K}\frac{B_{s,j}(k)}{A_{s,j}(k)}
		$}
\end{equation}
\begin{equation}
	\resizebox{\linewidth}{!}{$
		\frac{\partial \tilde{H}_{s,m}(k)}{\partial A_{s,\kappa}(k)}=-H_{s,m}(k)\cdot V_{s}\cdot
		\frac{B_{s,\kappa}(k)}{A_{s,\kappa}^2(k)}\prod_{j=1,j\neq \kappa}^{K}\frac{B_{s,j}(k)}{A_{s,j}(k)}
		$}
\end{equation}

The partial derivative of $\tilde{h}_{s,m}$ with respect to a generic parameter $p_{s,\kappa}$ is calculated exploiting the conversion from time to frequency and from frequency to time using again the Wirtinger calculus: 
\begin{equation}
	\resizebox{\linewidth}{!}{$
		\begin{split}
			\frac{\partial \tilde{h}_{s,m}}{\partial p_{s,\kappa}}=&	\Bigg(\sum_{n=0}^{N-1}cos\Big(2\pi\frac{kn}{N}\Big)+\sum_{n=0}^{N-1}sin\Big(2\pi\frac{kn}{N}\Big)\Bigg)\cdot\Bigg[\frac{\partial b_{0_{s,\kappa}}}{\partial p_{s,\kappa}}\cdot\sum_{k=0}^{N-1}\frac{\partial \tilde{H}_{s,m}(k)}{Re(B_{s,m}(k))} +\\&\frac{\partial b_{1_{s,\kappa}}}{\partial p_{s,\kappa}}\cdot\Big(\sum_{k=0}^{N-1}\frac{\partial \tilde{H}_{s,m}(k)}{Re(B_{s,m}(k))}cos\Big(2\pi\frac{k}{N}\Big)-\sum_{k=0}^{N-1}\frac{\partial \tilde{H}_{s,m}(k)}{Im(B_{s,m}(k))}sin\Big(2\pi\frac{k}{N}\Big)\Big)+\\&\frac{\partial b_{2_{s,\kappa}}}{\partial p_{s,\kappa}}\cdot\Big(\sum_{k=0}^{N-1}\frac{\partial \tilde{H}_{s,m}(k)}{Re(B_{s,m}(k))}cos\Big(2\pi\frac{2k}{N}\Big)-\sum_{k=0}^{N-1}\frac{\partial \tilde{H}_{s,m}(k)}{Im(B_{s,m}(k))}sin\Big(2\pi\frac{2k}{N}\Big)\Big)+\\&\frac{\partial a_{1_{s,\kappa}}}{\partial p_{s,\kappa}}\cdot\Big(\sum_{k=0}^{N-1}\frac{\partial \tilde{H}_{s,m}(k)}{Re(A_{s,m}(k))}cos\Big(2\pi\frac{k}{N}\Big)-\sum_{k=0}^{N-1}\frac{\partial \tilde{H}_{s,m}(k)}{Im(A_{s,m}(k))}sin\Big(2\pi\frac{k}{N}\Big)\Big)+\\&\frac{\partial a_{2_{s,\kappa}}}{\partial p_{s,\kappa}}\cdot\Big(\sum_{k=0}^{N-1}\frac{\partial \tilde{H}_{s,m}(k)}{Re(A_{s,m}(k))}cos\Big(2\pi\frac{2k}{N}\Big)-\sum_{k=0}^{N-1}\frac{\partial \tilde{H}_{s,m}(k)}{Im(A_{s,m}(k))}sin\Big(2\pi\frac{2k}{N}\Big)\Big)\Bigg]
		\end{split}
		$}
\end{equation}
where, through  Wirtinger calculus, we have that: $\frac{\partial B(k)}{\partial b_{0,\kappa}}=\frac{\partial A(k)}{\partial a_{0,\kappa}}=1 $
and $\frac{\partial B(k)}{\partial b_{1,\kappa}}=\frac{\partial B(k)}{\partial b_{2,\kappa}}=\frac{\partial A(k)}{\partial a_{1,\kappa}}=\frac{\partial A(k)}{\partial a_{2,\kappa}}=cos(2\pi\frac{k}{N})-sin(2\pi\frac{k}{N})$.

The local derivative of the IIR filter coefficients with respect to the parameters $f_{c}$, $V_{0,dB}$ and $Q$ can be calculated through the partial derivatives of the equations in \tableref{table:coeff_iir}. 
For example, the local derivative of the coefficient $b_{s,\kappa,0}$ with respect to $f_{c_{s,\kappa}}$ for the Boost filter is:

\begin{equation}
	\resizebox{\linewidth}{!}{$
	\frac{\partial b_{s,\kappa,0}}{\partial f_{c_{s,\kappa}}}=\frac{\pi\cdot \Big(\frac{1}{Q_{s,\kappa}}-\frac{10^{V_{s,\kappa,dB}/20}}{Q_{s,\kappa}}\Big)\cdot \Big(tan^2(\pi\frac{f_{c_{s,\kappa}}}{f_s})-1\Big)\cdot\frac{1}{cos^2(\pi\frac{f_{c_{s,\kappa}}}{f_s})}}{f_s\cdot \Big(\frac{1}{Q_{s,\kappa}}tan(\pi\frac{f_{c_{s,\kappa}}}{f_s})+tan^2(\pi\frac{f_{c_{s,\kappa}}}{f_s})+1\Big)^2}
	$
	}
\end{equation}

Finally, the derivative of the denormalization step is: 
\begin{equation}
	\frac{\partial q}{\partial p}=\frac{q_{max}-q_{min}}{2}
\end{equation}
where $p$ is the normalized parameter, $q$ is any of the equalizer parameters (denormalized), i.e. $f_c, Q, V_{0,dB}, V_{s,dB}$ and the terms $q_{max}, q_{min}$ denote its range. We must remind that for $f_c$ the range is different for each SOS (see Eq. \eqref{eq:fcdenorm}), while the other parameters have identical ranges for each SOS. 

The local derivative with respect to the regularization term  is given by the cascade of the following local derivatives:
\begin{equation}\label{eq_d_l2_d_y}
	\frac{\partial L_2}{\partial \tilde{h}_{s,m}(n)}=\frac{\partial L_2}{\partial \hat{r}_{s,m}}\cdot \frac{\partial \hat{r}_{s,m}}{\partial \hat{\epsilon}_{s,m}} \cdot \frac{\partial \hat{\epsilon}_{s,m}}{\partial \tilde{h}_{s,m}(n)}
\end{equation}
where:
\begin{equation}\label{eq:d_l_d_r}
	\frac{\partial L_2}{\partial \hat{r}_{s,m}}=\frac{\hat{r}_{s,m}-r_{s,m}}{\sqrt{\sum_{s=1}^{\mathcal{S}}(\hat{r}_{s,m}-r_{s,m})^2}},
\end{equation}
\begin{equation}\label{eq:d_r_d_e}
	\frac{\partial \hat{r}_{s,m}}{\partial \hat{\epsilon}_{s,m}}=-\frac{\hat{\epsilon}_{1,m}}{\hat{\epsilon}_{s,m}^2},
\end{equation}
\begin{equation}\label{eq:d_ep_d_y}
	\frac{\partial \hat{\epsilon}_{s,m}}{\partial \tilde{h}_{s,m}(n)}=2\cdot \tilde{h}_{s,m}(n).
\end{equation}

\section{Comparative Methods}\label{sec:baseline}

To conduct a thorough analysis of the proposed method we compare it to several additional techniques that can be employed for the room equalization task. We propose two DO techniques, both based on convolutional neural networks, an evolutionary IIR filter design method based on the Direct Search Method \cite{ramos2006filter} and the well-known Frequency Deconvolution method for the design of FIR filters \cite{kirkeby1998freq}.

\subsection{Alternative Deep Optimization Networks}

In addition to BiasNet, two DO architectures are proposed based on similar principles.
The first is a regular CNN, as in our previous work \cite{pepe2020designing}. This network has been adapted to the design of IIR filters as described with BiasNet, i.e. by generating the IIR parameters $f_c, Q, V_{0,dB}, V_{s,dB}$ and using the denormalization rule described in Eq. \eqref{eq:fcdenorm}. The network employs a fixed input, the RIR tensor, as done in our previous work.

The second network is a CNN with variable input. The idea stems from the same observations drawn in describing the BiasNet, i.e. that of a network adapting its input, in addition to its weights. At the first iteration we use the RIR as input tensor, however, at each next iteration the network is fed with the equalized RIR. This network, therefore, establishes an input-output feedback, since the equalized response generated at the $j$-th iteration is fed as input at iteration $(j+1)$. In the following, we shall call this architecture Convolutional Feedback Network (CFN).

The rationale for testing the CNN is verifying how well it performs for the new task, after adapting its output layers for the IIR equalizer parameters design, and assessing the computational cost and performance gain of the proposed BiasNet architecture. The CFN, on the other hand, provides an alternative approach to input optimization, that is worth investigating in terms of performance. Similarly to BiasNet it is able to learn an optimal input, however it uses a different strategy and a larger number of parameters.

\subsection{Direct Search Method}

The Direct Search Method is a heuristic technique that is used for optimization problems \cite{Hooke1961}. This method is a derivative-free optimization algorithm and can work with no constraints. As will be described below, this technique is simple to implement and has been used to solve many optimization problems \cite{LEWIS2000191}.

The authors in \cite{ramos2006filter} implemented the Direct Search Method for an IIR parametric equalizer. This heuristic algorithm is described as in \cite{behrends2011automatic}: a random variation $\Gamma$ in the range $-\gamma\leq \Gamma \geq \gamma$ is added to a generic parameter vector $c$:

\begin{equation}
	\hat{c}_i=c_i\cdot(1+\Gamma)
\end{equation}

If the new parameters achieve a better cost function, they are kept, otherwise they are rejected and another random variation $\Gamma$ is performed. The process continues until the demands have been met. For our case, $c$ is composed of the parameters of the IIR filters $f_c, Q, V_{0,dB}, V_{s,dB}$.

\subsection{Frequency Deconvolution}

This method is based on deconvolution in the frequency domain to estimate a room inverse that is optimal in the Least Squares sense \cite{kirkeby1998freq}, i.e. the cost function is convex. The optimal filters $G(k)$ are computed in the frequency domain by:
\begin{equation}
	G(k)=[H^{H}(k)H(k)+\beta_{FD} I]^{-1}H^{H}(k)D(k)
\end{equation}
where $I$ is the identity matrix, $^H$ is the Hermitian operator, $H(k)$ is the transfer function matrix of the impulse responses, $D(k)$ is the target frequency responses matrix and $\beta_{FD}$ is a regularization term that avoids extreme peaks in the inverse filters. The inverse FFT and then a circular shift of K are computed, where K is the FFT size, to achieve the filters in the time-domain. Conducting deconvolution in the frequency domain speeds up the computation, by taking advantage of the FFT algorithm. 

This method is generally taken as a reference baseline for FIR equalization of a room impulse response and its solution is regarded as optimal. It must be noted that the solution is such only in the Least Squares sense, therefore by posing the problem in non-convex terms, as is done with our proposed method, other solutions may exist that solve the room equalization problem with similar or superior spectral flatness.

\section{Experiments}\label{sec:exp}

\begin{figure}[tbp]
	\centering
		\includegraphics[width=0.5\linewidth]{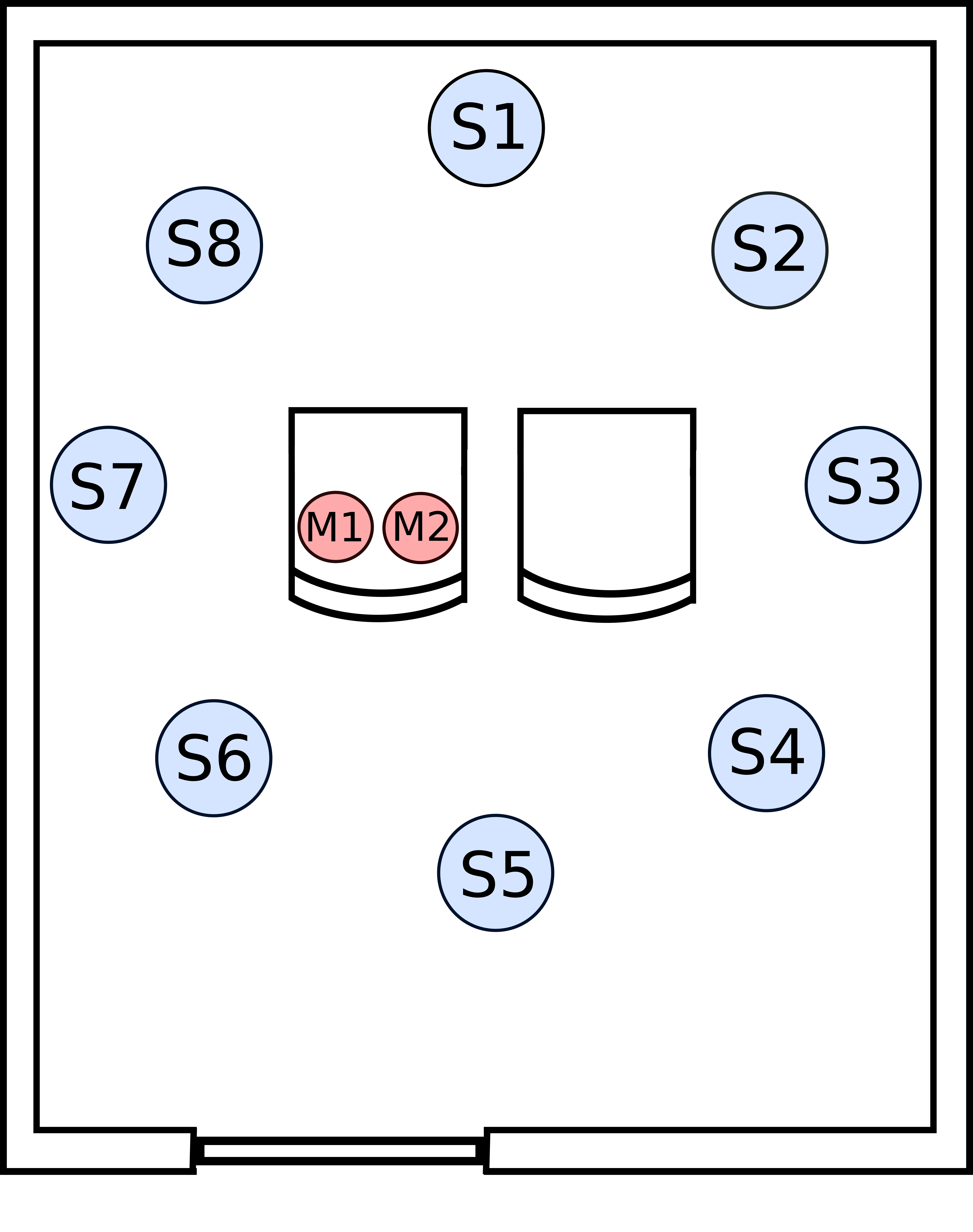}
	\caption{Top view of the room  showing the placement of the speakers and microphones.}
	\label{fig:room_scheme}
\end{figure}
\begin{figure}[tbp]
	\centering
	\includegraphics[width=\linewidth]{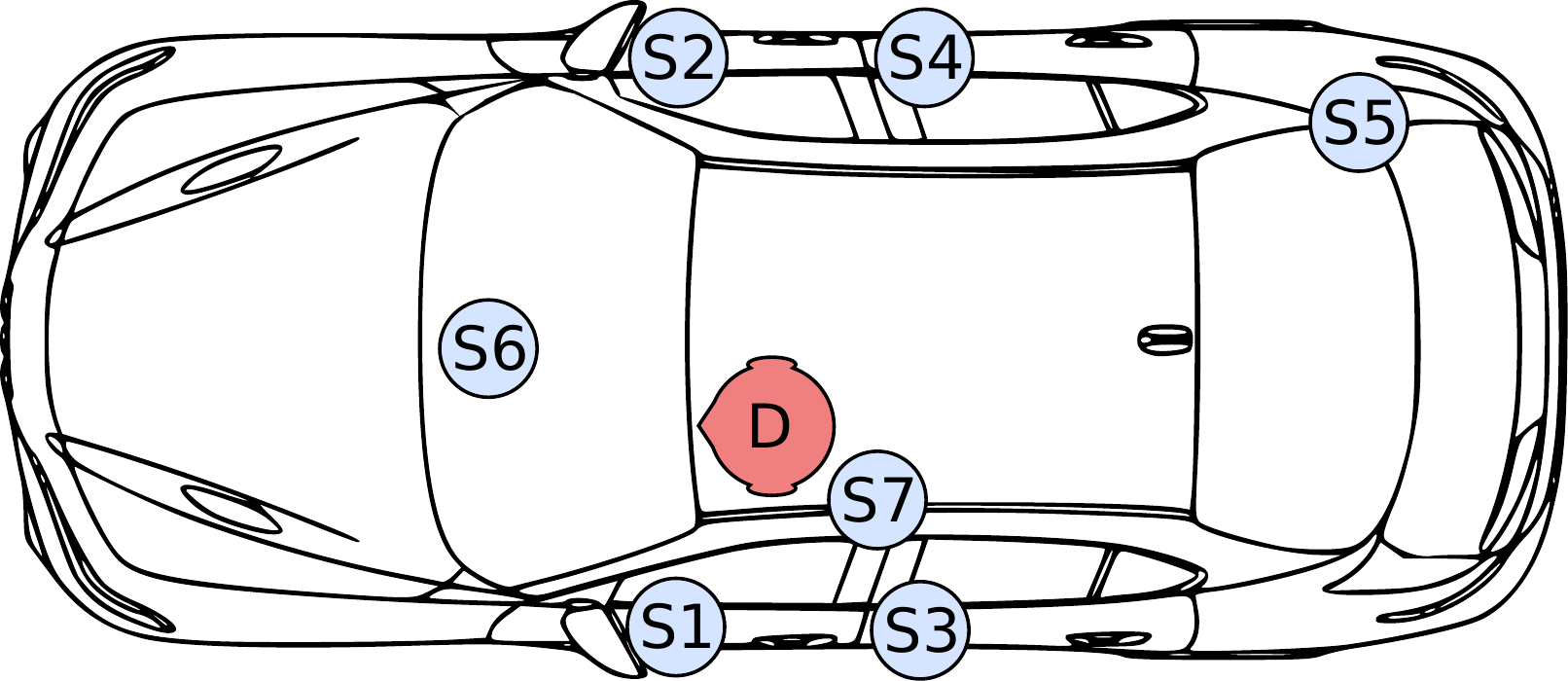}
	\caption{Top view of the car's cabin showing the placement of the speakers and the microphones. D corresponds to the mannequin head with microphones mounted in the ears.}
	\label{fig:giulia_scheme}
\end{figure}

To validate the proposed method, we perform experiments with two scenarios: a regular room and a car cabin.
The first scenario is a rectangular room of dimension $4.0 \times 5.5 \times 3.0$\,m and composed of 8 speakers circularly placed around two seats, as shown in \figref{fig:room_scheme}. The left seat is fitted with two omnidirectional microphones simulating the listener's ears. The loudspeakers are mid-woofers with a frequency range between 100\,Hz and 15\,kHz. Impulse responses were measured using the exponential sine sweep method \cite{farina2007advancements} with a sampling frequency of 48\,kHz, using a RME Madiface audio interface, and a Dante-equipped amplification system.

The second scenario is a car cabin, an Alfa Romeo Giulia, the same used for our previous works. Impulse responses were measured using the sine sweep method \cite{farina2007advancements} and measuring them using Aurora plug-in\footnote{http://pcfarina.eng.unipr.it/Aurora\_XP/index.htm} and a Kemar 45BA mannequin placed on the driver's seat. The sampling frequency equals 28.8\,kHz and then oversampled to 48\,kHz.

The two scenarios exhibit different characteristics. The room employs only one type of loudspeakers arranged in a regular pattern. These cover slightly more than two decades of the audio range and the maximum excursion between minima and maxima in the unequalized frequency responses never exceed $\sim 15$\,dB. On the contrary, the car is fitted with loudspeakers of different size and bandwidth, arranged in a irregular fashion. Furthermore, the car material absorption coefficient varies largely. This results in a large excursion of the frequency response ($>$20\,dB) and a wider range to equalize, covering almost three decades. The car cabin is, therefore, a more challenging scenario.

Preliminary experiments were, thus, conducted in the room, to gather more insights on the described equalization techniques, to compare the DO methods discussed above and find suitable hyperparameters. More specifically, we conducted a hyperparameters search in the MIMO case and kept the best architecture for all subsequent experiments. The experiments include SISO, MIMO and multiple-input single-output (MISO) cases for the room environment.
After discussing these results we turn our attention to the car cabin scenario, using all its speakers and microphones ($7 \times 2$ MIMO scenario).

The desired frequency response for our work is a 0\,dB flat band in the range 100\,Hz-14\,kHz for the room scenario, and 20\,Hz-14\,kHz for the car scenario. 

Both optimization and evaluation are conducted using a third-octave band averaging for the frequency responses. This choice is motivated by the human ear resolution which is not very sensitive to narrow dips and notches. The number of SOS's (and, similarly, the bands used during denormalization of the $f_c$ parameter) is equal to the number of third-octave bands within the speaker's operating frequency range. This makes for 22 SOS's for each speaker in the room scenario, and from 21 to 29 in the cabin scenario. Therefore, the number of parameters to optimize is 536 and 529, respectively. The ranges of the other parameters, identical for all SOS's, are: $Q_{min}=0.05$, $Q_{max}=5.0$, $V_{0,min,dB}=-10$\,dB, $V_{0,max,dB}=10$\,dB, $V_{s,min,dB}=-20$\,dB, $V_{s,max,dB}=20$\,dB.

Regarding the FD, we searched for the best regularization term $\beta_{FD}$ following the strategy used in \cite{pepe2020designing} and found that a constant $\beta_{FD} = 10^{-4}$ works well for all the experiments. Finally, for the DSM we set $\gamma = 0.01$.

The performance is evaluated using the averaged Mean Square Error ($\overline{MSE}$) and the standard deviation ($\overline{\sigma}$) \cite{pepe2020designing} in the third-octave band and within the desired frequency range.

\begin{equation}
	\label{eq:mse}
	\overline{MSE}=\frac{1}{\mathcal{M}}\sum_{m=1}^{\mathcal{M}}\Bigg(\frac{\sum_{\omega=\omega_l}^{\omega_h}\Big(|\tilde{H}_{m,1/3}(\omega)|-|H_{des,1/3}(\omega)|\Big)^2}{\omega_h-\omega_l}\Bigg)
\end{equation}

The average standard deviation $\overline{\sigma}$ is calculated as:
\begin{equation}
	\overline{\sigma}=\frac{1}{\mathcal{M}}{\sum_{m=1}^{\mathcal{M}}\sigma_m}
\end{equation}
where $\sigma_m$ is the standard deviation of $m$-th microphone:
\begin{equation}
	\sigma_m=\sqrt{\frac{1}{\omega_h-\omega_l+1}\sum_{\omega=\omega_l}^{\omega_h}(10\cdot log_{10}|\tilde{H}_{m,1/3}(\omega)|-D)^2}
\end{equation}
\begin{equation}
	D=\frac{1}{\omega_h-\omega_l+1}\sum_{\omega=\omega_l}^{\omega_h}(10\cdot log_{10}|\tilde{H}_{m,1/3}(\omega)|)
\end{equation}

Before performing the optimization, pre-processing is performed. The delay of each speaker is automatically determined by finding the instant in which its direct sound reaches the reference microphone. Then we determine the offset gain to be added in the optimization, so as to normalize the output frequency responses to 0\,dB.

The number and dimension of the layers of a neural network impact significantly the performance. Therefore, a preliminary random hyperparameters search has been conducted to find a suitable network architectures that have been, thus, retained for all the experiments. In that regard, the BiasNet has two degrees of freedom: the number of layers and their size, while the CFN and the CNN have more degrees of freedom: the convolutional layer number, horizontal and vertical size and pooling, the fully connected layer number and size. The activation function used in the network was found to be a less sensitive parameter. We adopted the sine activation function, which was recently proposed in \cite{sitzmann2020implicit} and shows to be well suited for optimization as it avoids local minima during network optimization. This activation function, behaves well for backpropagation as its derivatives do not vanish.

We tested 30 BiasNet configurations varying the number of layers from 1 to 10 and the number of neurons from 16 to 4096. 
For the CNN and CFN we tested 1 or 2 convolutional layers: the number of kernels equals 25 in the case of 1 convolutional layer,  48 and 24 or 100 and 10 in the case of 2 layers, respectively. The dimension of the kernel for the first and second convolutional layer equals $\mathcal{M}\times 1$ and $1 \times \mathcal{S}$, respectively. The number of hidden layers varied from 1 to 4, with the number of neurons ranging from 32 to 1024.

In all the experiments the Adam optimizer \cite{kingma2014method}, was employed using a learning rate of $10^{-4}$, and parameters $\beta_1 = 0.9, \beta_2 = 0.999$. The number of iterations was set to 10,000 and the weights and bias have been initialized with a uniform distribution.

Deep optimization methods were implemented in \textit{Python} using \textit{Tensorflow}\footnote{www.tensorflow.org} 2.0.0, while DSM and FD were implemented in \textit{Matlab}. The experiments were performed using a machine with an Intel i7 processor, 32\,GB of RAM and an Nvidia Titan GPU with 12\,GB of dedicated RAM.

\section{Results} 
\label{sec:results}

\subsection{Neural Architectures Comparison}

We first consider the three neural networks (BiasNet, CFN and CNN) for equalization in the room scenario. We conduct experiments to find the best network hyperparameters, giving priority to the MIMO case. The same networks will be tested for SISO and MISO cases without a further hyperparameters search, for the sake of conciseness. Results are reported in Table \ref{tab:preliminary}. As can be seen, many architectures in the test provide similar performance ($\overline{MSE} \sim 10^{-5}$), however the difference in the number of trainable parameters is extremely large. Taking the CNN as a reference, the best CFN achieves slightly better performance at the cost of an order of magnitude more parameters. On the other hand, the BiasNet excels in the test with a significantly lower number of parameters. Configuration \#5, has only 17,664 trainable parameters, but achieves almost identical performance as configuration \#26 which is 40 times larger to train. However, there is a critical number of trainable parameters under which the performance decreases: configuration \#9 is similar to \#26, besides having half the neurons in its only layer, but closely follows the performance of \#26. On the contrary, progressively halving the neurons of the only layer (see configurations \#8 and \#10) makes the performance fall. The worst case is achieved with configuration \#4 which has only the output layer, with learnable bias, and results in an unacceptable equalization performance.

\begin{table}[tb]
	\centering
	\caption{Preliminary test comparing several neural networks in the MIMO configuration for the room scenario. The number of neurons for each hidden layer is shown in round brackets in the Layers column.}
	\label{tab:preliminary}
	\begin{tabular}{r||c||l||l}
		\hline
 		Architecture	& $\overline{MSE}$ & Layers & \begin{tabular}[c]{@{}l@{}}No. Learnable \\ Parameters\end{tabular} \\
 			\hline
		BiasNet \#26 & $1.18 \cdot 10^{-5}$ & (1024, 512, 256, 128) & 758,784 \\
		BiasNet \#5 & $1.19 \cdot 10^{-5}$ & (256) & 137,728 \\
		BiasNet \#9 & $1.29 \cdot 10^{-5}$ & (128) & 68,864 \\
		CFN & $1.38 \cdot 10^{-5}$ & best CFN\footnotemark &4,673,390\\
		BiasNet \#7  & $1.46 \cdot 10^{-5}$   & (256, 256, 256)  & 268,800 \\
		CNN &  $1.66 \cdot 10^{-5}$ & same as \cite{pepe2020designing} & 2,369,390 \\
		BiasNet \#25 & $1.68 \cdot 10^{-4}$ & \begin{tabular}[c]{@{}l@{}}(16,32,64,128,256,512,\\ 256,128,64,32,16)\end{tabular}  & 357,792 \\
		BiasNet \#8 & $4.24 \cdot 10^{-4}$   & (64) &  34,432   \\
		BiasNet \#10 & $6.36 \cdot 10^{-4}$ & (32) & 17,216 \\
		BiasNet \#4 & $9.56 \cdot 10^{-1}$ & out only (with bias) & 536\\
		No EQ & $0.377$ & - & - \\
		\hline
	\end{tabular}
\end{table}

\footnotetext{Note: the CFN was the best among all the tested CFN and is composed of 2 convolutional layer of 100 and 10 kernels, respectively, 3 dense layers of 64 neurons each.}

\begin{figure}[tb]
	\centering

	\includegraphics[width=\linewidth,trim={2cm 10cm 2cm 10cm},clip]{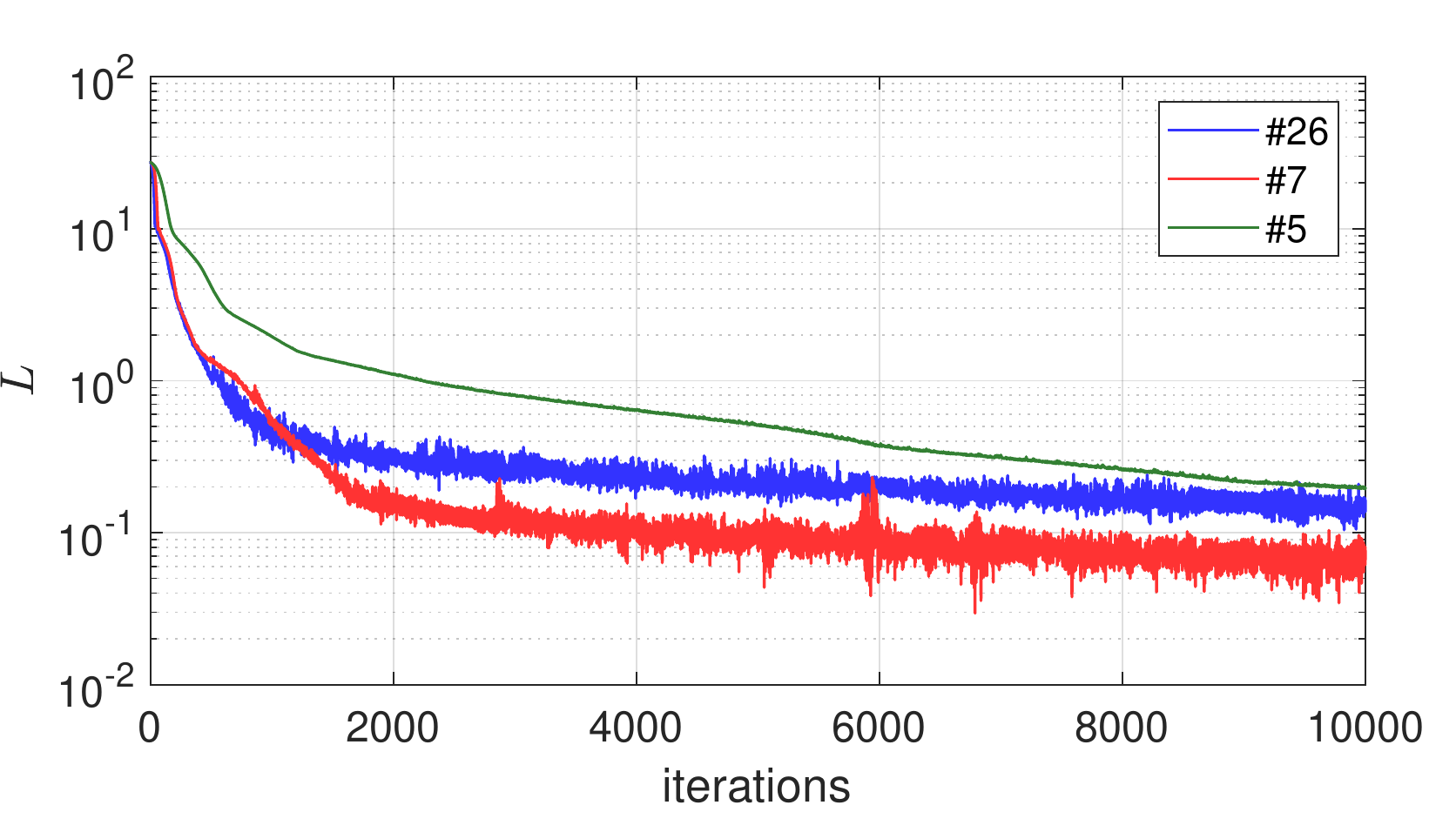}

	\caption{Minimization of the loss function for configurations \#26 (blue line), \#7 (red line), \#5 (green line) for MIMO equalization in the room scenario. The optimization speedup is directly related to the number of trainable parameters. It must be noted that a lower loss does not mean a lower result as the loss includes the regularization term and is computed as the root of squares, while the MSE is computed as the average of squared terms.}
	\label{fig:loss}
\end{figure}

Although several networks obtain almost identical performance in sheer terms of performance, it is worth considering the optimization time achieved by networks with very different number of trainable parameters. The optimization time depends on two factors: the number of iterations to reach a target goal and the time required by each iteration. The latter, in our case, mostly depends on: \textit{(a)} the time required for computing of the neural network graph and its backpropagation; \textit{(b)} the time required for computing the filter coefficients, and simulating the room. When neural networks run on a GPU, the latter can be quite expensive, especially for the MIMO case, and adds a fixed cost that is not negligible, constraining the overall optimization time to be of the order of several minutes, with the current hardware and software implementation. For this reason, it is worth aiming at a reduction of the number of iterations to reach a desired goal (a target $\overline{MSE}$ or a substantial stop in the loss descent). We trace the loss descent during optimization, plotted in \figref{fig:loss}. As we can see, with BiasNet, the number of parameters directly relates to a speedup of the optimization process. 

For the reasons outlined above among the tested networks we select the first in Table \ref{tab:preliminary}, that provides the best performance and fast optimization times.

\subsection{Room Scenario} \label{sec:room_res}

This section describes results for SISO, MISO and MIMO equalization in the room scenario, considering the BiasNet and the non-neural comparative methods. The SISO results are presented in \tableref{table:room_siso}, showing that the neural approach achieves better results than the baseline techniques by one or more orders of magnitude, without the need for refining the neural network size and hyperparameters. It is worth noting that the DSM is unable to improve much on the non-equalized case given the same number of iterations of the DO approach. The FD approach achieves excellent performance, however its (convex) optimum does not coincide with the (non-convex) optimal solution found by the BiasNet, which is superior. Furthermore, the computational cost of a 8192-th order FIR filter is very high.

\begin{table}[tb]
		\renewcommand{\arraystretch}{1.3}
	\caption{Results for SISO equalization, room scenario.}
		\label{table:room_siso}
			\centering
	\begin{tabular}{c||c||c}
		\hline
		Method&${MSE}$&${\sigma}$\\\hline
		BiasNet&$\bf{1.32\cdot 10^{-5}}$	&$\bf{1.58\cdot 10^{-2}}$\\
		FD$_{1024}$&$9.74\cdot 10^{-3}$	&$4.36\cdot 10^{-1}$	\\
		FD$_{8192}$&$2.08\cdot 10^{-4}$&$5.54\cdot 10^{-2}$\\
		DSM&$3.43\cdot 10^{-2}$	&	$9.92\cdot 10^{-1}$\\
		No EQ & $3.80 \cdot 10^{-1}$ & $1.69$ \\\hline
	\end{tabular}

\end{table}

MISO results are presented in Table \ref{table:room_miso}. In this table we report the results at the listening point used for optimization (right microphone), as well as the average of the $\overline{MSE}$ calculated at both listening positions. As can be seen, the performance obtained in the former case is extremely high, showing that an increase in the number of loudspeakers improves the equalization performance with respect to the SISO case. However, considering only one listening position for optimization provides a solution that does not work well for the other listening position, therefore the performance of the best methods in the table drops by at least 4 orders of magnitude when evaluating the performance at both microphones. It is worth noting that in this case the FD$_{1024}$ method is inferior to the IIR filters provided by DSM, while the FD$_{8192}$ slightly improves over the proposed method. However, the solution found by FD$_{8192}$ suffers a more sever degradation of performance when the error is computed at both listening positions.

Finally, Table \ref{table:room_mimo} shows that the proposed method scores optimally. All methods fail to achieve results as good as those computed for the MISO case (right microphone), since the computed solution must fit two listening positions. This is expected, since rising the number of microphones $\mathcal{M}$ increases the complexity of the problem.

\begin{table}[tb]
	\renewcommand{\arraystretch}{1.3}
	
	\caption{Results for MISO equalization, room scenario evaluated at the listening point used during optimization (Right Mic) and both microphones (L+R Mic).}
	\label{table:room_miso}
	\centering
	\resizebox{\linewidth}{!}{
		\begin{tabular}{c||c||c||c||c}
			\hline
			\multirow{2}{*}{Method}&\multicolumn{2}{c||}{Right Mic}&\multicolumn{2}{c}{L+R Mic}\\\cline{2-5}\cline{2-5}
			&${MSE}$&${\sigma}$&$\overline{MSE}$&$\overline{\sigma}$\\\hline
			BiasNet &$8.95\cdot 10^{-6}$	&$1.26\cdot 10^{-2}$	&$\mathbf{5.53\cdot 10^{-2}}$	&$\mathbf{8.04\cdot 10^{-1}}$	\\
			DSM&$1.87\cdot 10^{-2}$	&$6.98\cdot 10^{-1}$	&$8.76\cdot 10^{-2}$	&1.42	\\
			FD$_{1024}$&$5.26\cdot 10^{-2}$	&$3.55\cdot 10^{-1}$	&$1.63\cdot 10^{-1}$	&$9.05\cdot 10^{-1}$	\\
			FD$_{8192}$&$\mathbf{4.88\cdot 10^{-6}}$	&$\mathbf{9.58\cdot 10^{-3}}$	&$1.30\cdot 10^{-1}$	&$9.52\cdot 10^{-1}$	\\
			No EQ &$3.92 \cdot 10^{-1}$ &$1.91$&$3.77 \cdot 10^{-1}$&$1.99$ \\
			\hline
		\end{tabular}
	}
\end{table}

\begin{figure*}
	\centering
	\begin{subfigure}[b]{0.45\linewidth}
		\includegraphics[width=\linewidth,trim={3.5cm 9cm 3cm 9.5cm},clip]{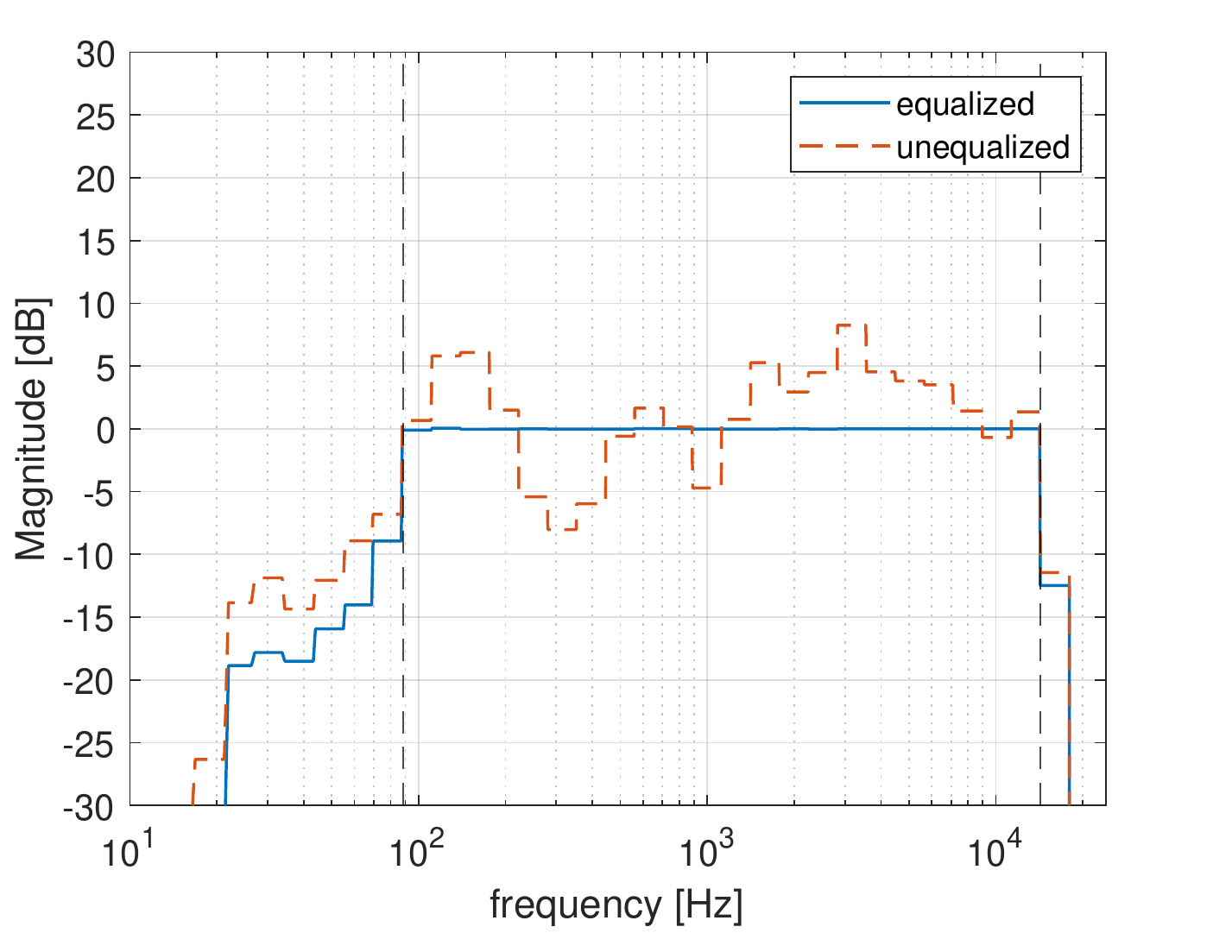}
		\caption{Left microphone}
	\end{subfigure}
	\begin{subfigure}[b]{0.45\linewidth}
		\includegraphics[width=\linewidth,trim={3.5cm 9cm 3cm 9.5cm},clip]{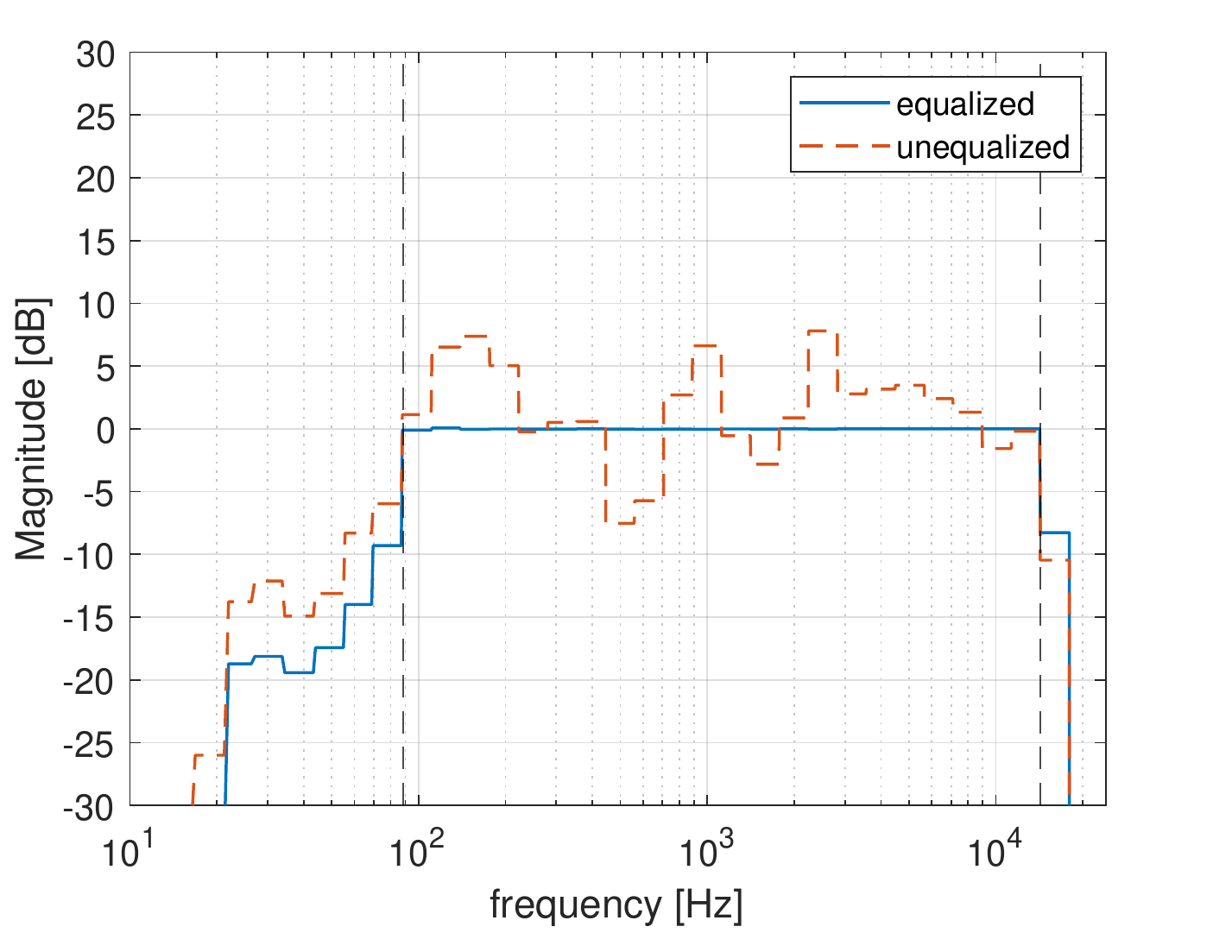}
		\caption{Right microphone}
	\end{subfigure}
	\caption{One third octave band magnitude response of (a) left and (b) right microphone in the room scenario. Red line is the unequalized frequency response, the blue line is the equalized one and the black dotted lines refer to the minimum and maximum frequency to be equalized}
	\label{fig:freq_room}
\end{figure*}

As an example of the frequency response that these method achieve, we show some of the magnitude responses in \figref{fig:freq_room}. The difference between the unequalized response (red line) and the equalized response (blue line) is evident. While the unequalized spectrum exhibits an excursion of more than 10\,dB, the equalized response is flat in the frequency range covered by the loudspeakers (vertical dashed lines).

\begin{figure}[tb]
	\centering

	\begin{subfigure}[b]{0.45\linewidth}
	\includegraphics[width=\linewidth,trim={3cm 7cm 5.25cm 7.5cm},clip]{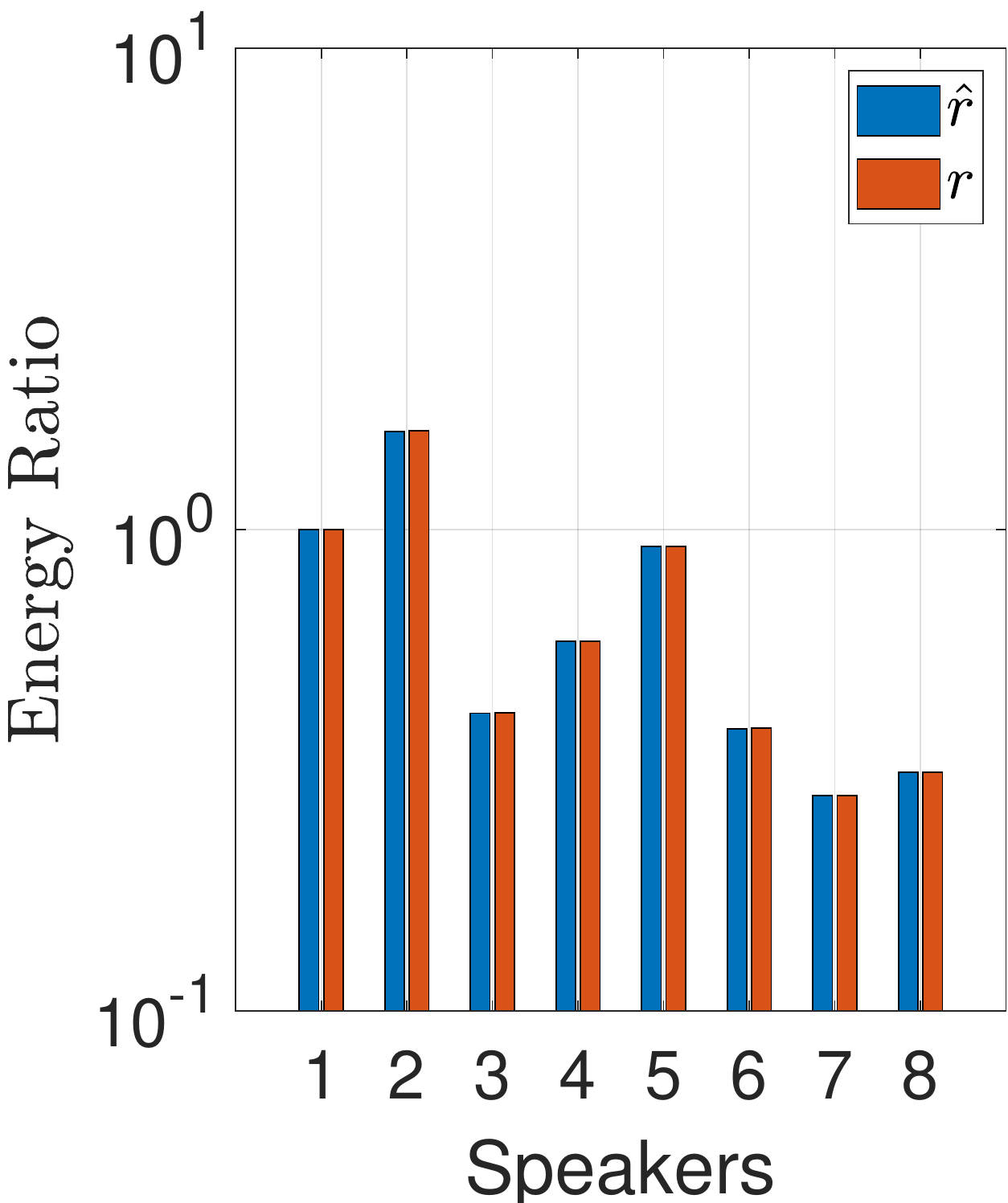}
		\caption{}
	\end{subfigure}
	\begin{subfigure}[b]{0.45\linewidth}
	\includegraphics[width=\linewidth,trim={3cm 7cm 5.25cm 7.5cm},clip]{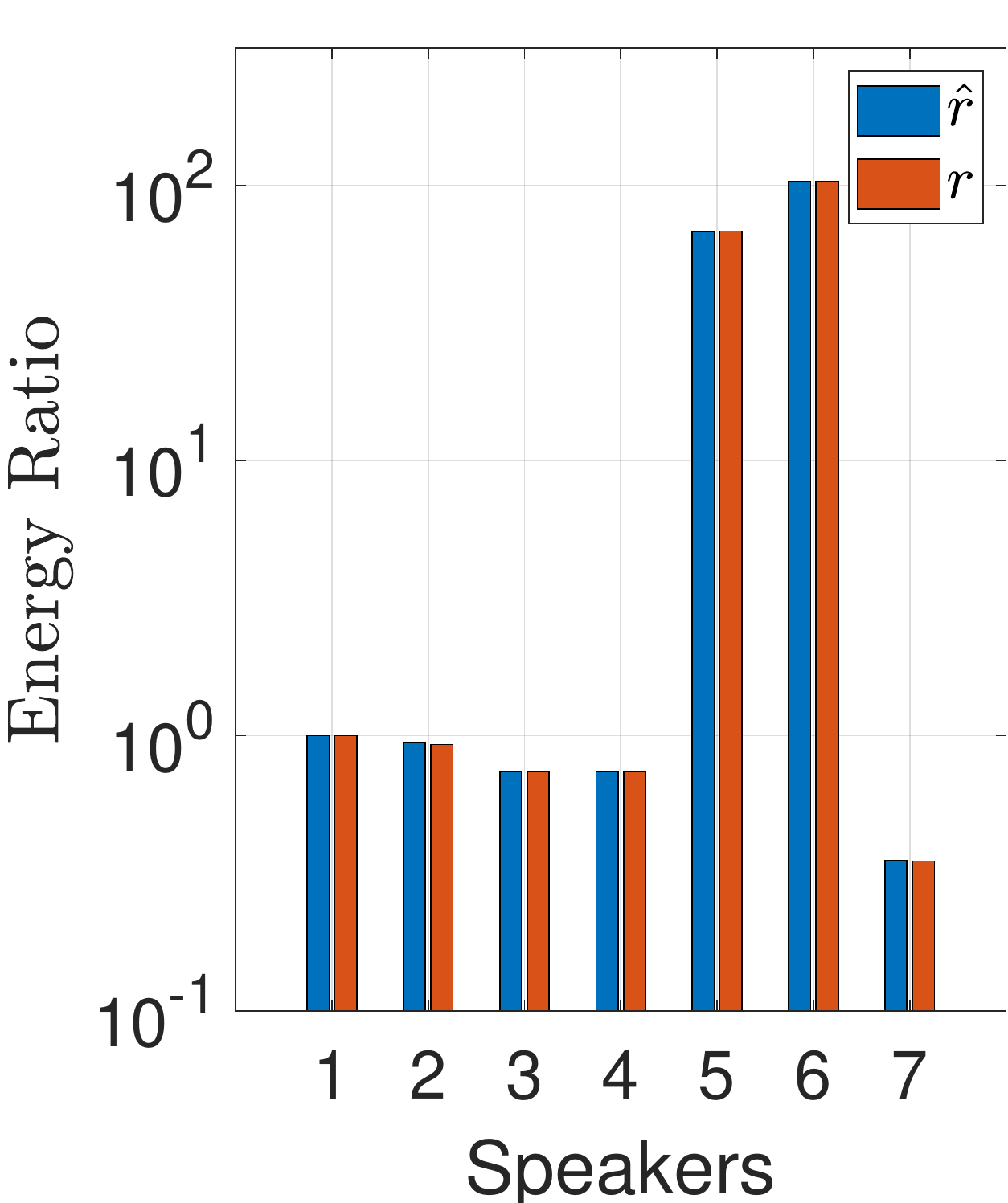}
		\caption{}
	\end{subfigure}
	\caption{Bar graph of energy ratio after ($\hat{r}$) and before $r$ the optimization for the room scenario (a) and the car scenario (b). Speakers 5 and 6 in (b) are woofer and subwoofer, therefore have larger energy.}
	\label{fig:ratio}
\end{figure}

As discussed previously, the equalization process should provide the desired spectral profile without altering the energy provided by the individual speakers. \figref{fig:ratio} shows that the energy of each speaker is preserved before and after the equalization process.

\begin{table}[tb]
	\renewcommand{\arraystretch}{1.3}
	
	\caption{Results for MIMO equalization, room scenario.}
	\label{table:room_mimo}
	\centering
	\begin{tabular}{c||c||c}
		\hline
		Method&$\overline{MSE}$&$\overline{\sigma}$\\\hline
		BiasNet &$\bf{1.18\cdot 10^{-5}}$	&$\bf{1.40\cdot 10^{-2}}$	\\
		DSM&$2.54\cdot 10^{-2}$	&$7.18\cdot 10^{-1}$	\\
		FD$_{1024}$&$4.57\cdot 10^{-2}$		&$4.36\cdot 10^{-1}$	\\
		FD$_{8192}$&$1.38\cdot 10^{-5}$		&$1.57\cdot 10^{-2}$	\\
		No EQ & $3.77 \cdot 10^{-1}$&$1.99$ \\
		\hline
	\end{tabular}
\end{table}

\begin{figure*}
	\centering
	\begin{subfigure}[b]{0.45\linewidth}
		\includegraphics[width=\linewidth,trim={3cm 9cm 3cm 9.5cm},clip]{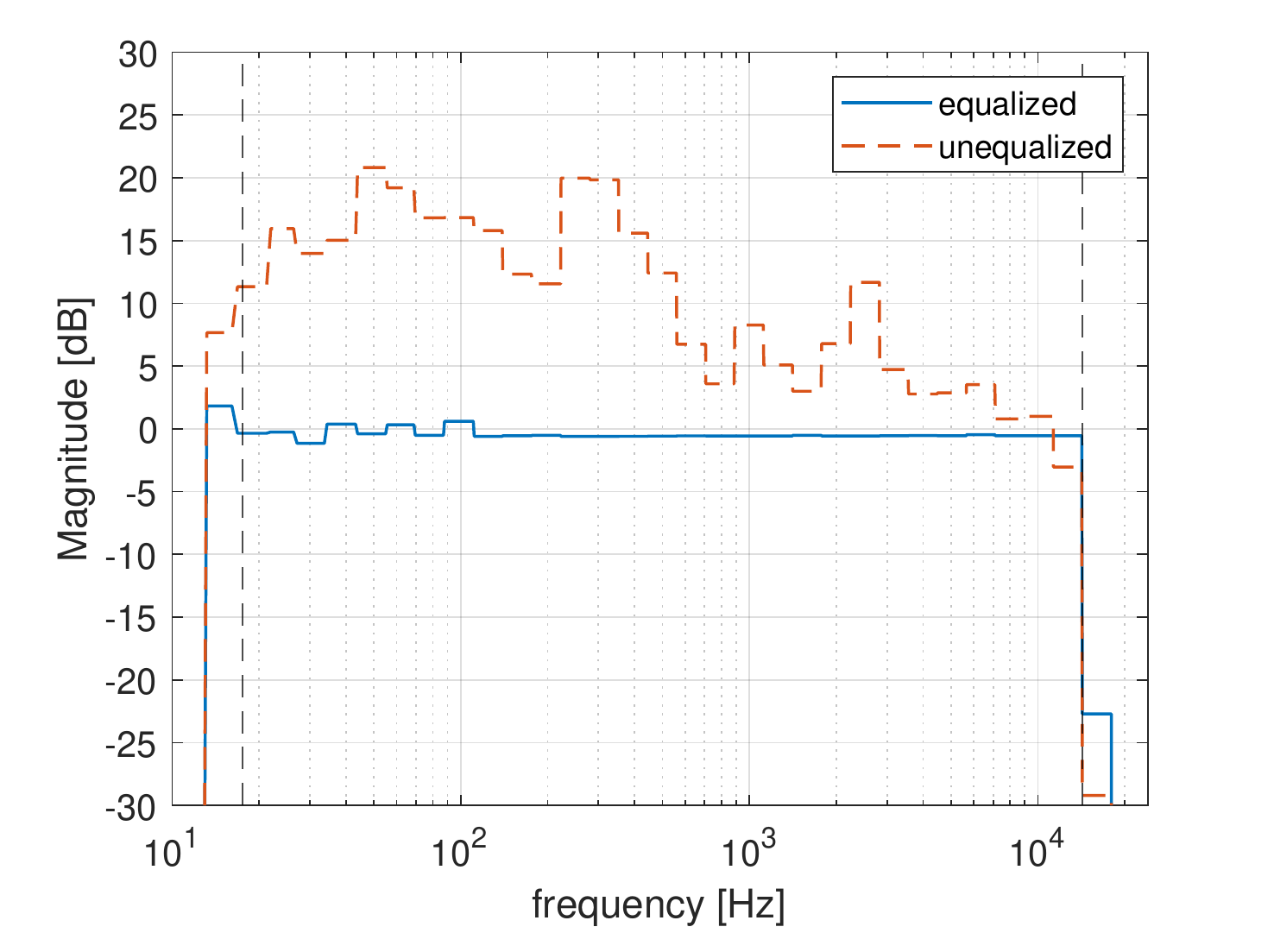}
		\caption{Left microphone}
	\end{subfigure}
	\begin{subfigure}[b]{0.45\linewidth}
		\includegraphics[width=\linewidth,trim={3cm 9cm 3cm 9.5cm},clip]{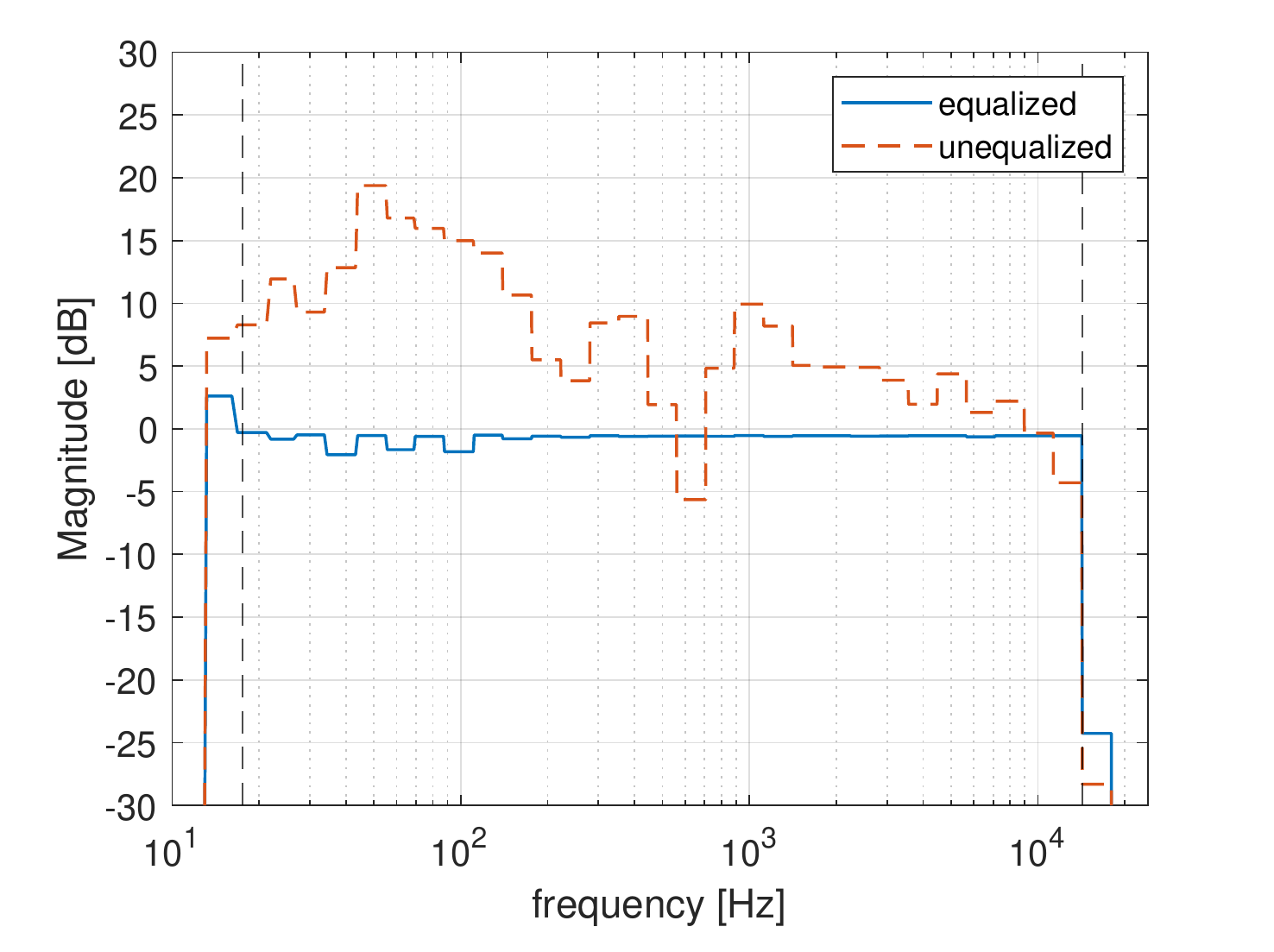}
		\caption{Right microphone}
	\end{subfigure}
	\caption{One third-octave band magnitude response of the measured signal at the reference microphones: (a) left and (b) right microphone in the car's cabin scenario. The vertical black dotted lines denotes the frequency range to be equalized.}
	\label{fig:freq_giulia}
\end{figure*}

\subsection{Car Scenario}

Results for the car cabin scenario are reported in \tableref{table:res_giulia}. Since the car provides a more challenging scenario, the $\overline{MSE}$ are higher than that of the room and the DSM fails to provide a decent equalization performance. The FD method does not match the performance achieved by the proposed method, which as in the previous experiments obtains the best results.

\begin{table}[tb]
	\renewcommand{\arraystretch}{1.3}
	\caption{Results for MIMO equalizatiom, car cabin scenario.}
	\label{table:res_giulia}
	\centering
	\begin{tabular}{c||c||c}
		\hline
		Method&$\overline{MSE}$&$\overline{\sigma}$\\\hline
		BiasNet &$\bf{5.74\cdot 10^{-3}}$	&$\bf{1.83\cdot 10^{-1}}$\\
		DSM	&3.62	&2.76	\\
		FD$_{1024}$&$4.22\cdot 10^{-2}$	&$8.15\cdot 10^{-1}$	\\
		FD$_{8192}$&$1.84\cdot 10^{-2}$	&$5.02\cdot 10^{-1}$	\\
		No Eq & $13.47$ & $3.16$ \\
		\hline
	\end{tabular}
\end{table}

In \figref{fig:freq_giulia} one third octave band frequency responses are presented: at high frequencies the amplitude responses are flat, both for the overlapping of the loudspeakers frequency responses and because the network optimizes in this range only two loudspeakers out of the seven installed in the car. At low frequencies the network has not been able to optimize as well as at high frequencies, indeed it presents a maximum deviation of 5\,dB around 40\,Hz.

This scenario makes a more challenging testbed for our method: the loudspeakers are not arranged in a regular pattern, their frequency ranges are different and the internal cabin volume is irregular. Despite these difficulties, the equalization is superior to the FD method, which is usually considered the optimal method for room equalization, and the energy of the loudspeakers signals is preserved.

\section{Remarks}\label{sec:remarks}

Our experiments show that the proposed method is comparable or slightly better than the FD$_{8192}$ method in terms of performance, however, the computational cost and the phase properties of the two differ. 

\begin{figure*}[tb]
	\centering
	\begin{subfigure}[b]{0.45\linewidth}
		\includegraphics[width=\linewidth,trim={0.5cm 2.8cm 1cm 0.5cm},clip]{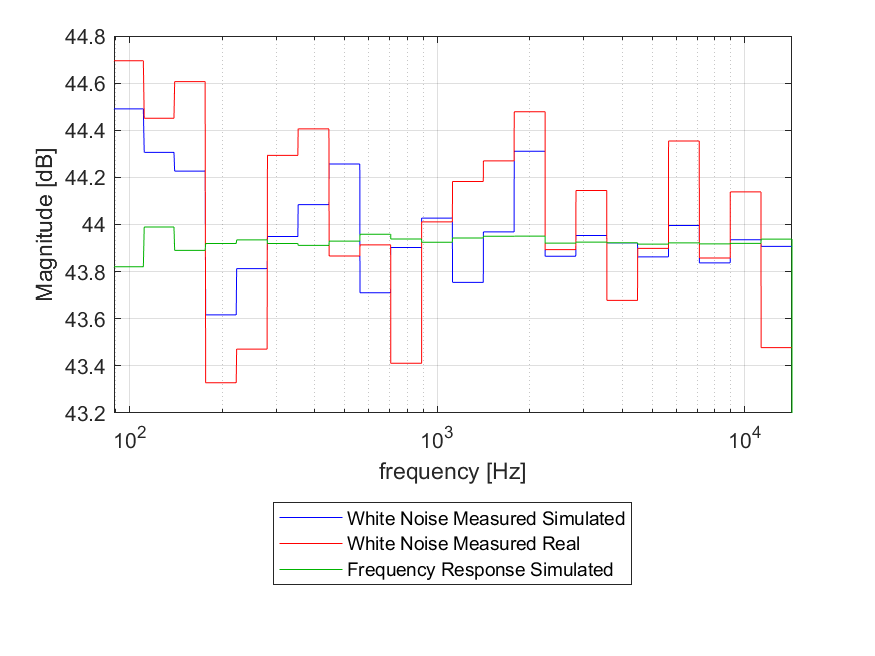}
		\caption{Left microphone}
	\end{subfigure}
	\begin{subfigure}[b]{0.45\linewidth}
		\includegraphics[width=\linewidth,trim={0.5cm 2.8cm 1cm 0.5cm},clip]{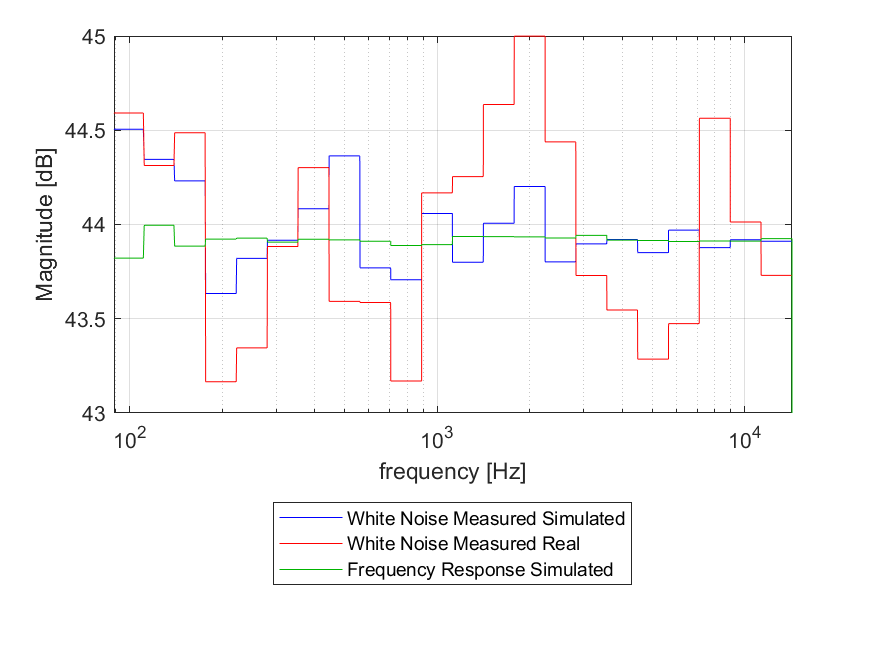}
		\caption{Right microphone}
	\end{subfigure}
	\caption{One-third octave band magnitude response of the measured signal at the reference microphones: (a) left and (b) right microphone in the room scenario. The green line is the equalized magnitude response depicted in \figref{fig:freq_room} which is computed by filtering a discrete-time impulse and transforming to the time domain. The blue line is simulated using white noise as input, while the red line is measured in the real scenario using white noise. Please note: the magnitude range is only 2 \,dB to emphasize the small differences.}
	\label{fig:freq:real}
\end{figure*}

\subsection{Computational Cost}

The computational cost of each of the $\mathcal{S}$ equalizers is 16383 floating point operations per sample in the FD$_{8192}$ case. The IIR equalizers are composed of 22 SOS's in the room scenario and 29 SOS's in the car scenario, totalling 198 and 261 operations per sample respectively, which means a reduction of the cost of nearly two orders of magnitude.
The cost grows further with an increase of the loudspeakers, e.g. in the MIMO room scenario the proposed method requires 1584 operations against 131064 operations for the FD$_{8192}$.

Clearly, it must be remarked that the FD method allows to design shorter filters. With FIR filters of order 1024 the performance is significantly lower but still acceptable, yielding an improvement of approximately one order of magnitude with respect to the non equalized case. However, the number of operations per filter is still significantly larger than the proposed method (2047 operations per sample).


\subsection{Real-World Validation}\label{sec:real_sim}

The evaluations provided above are referred to a simulated room, employing the RIRs. However, it is worth validating the experiments in the real scenario, to assess whether the simulations are accurate enough and the results are reproducible in a real environment.

The IIR filters designed for the room scenario in the MIMO case are taken to the real room and applied for equalization. More specifically, we load the designed IIR filters on a  Simulink\footnote{https://www.mathworks.com/products/simulink.html} patch to preprocess the signal and feed the loudspeakers. The hardware setup is the same as described in \secref{sec:exp}. We measure the frequency response of the environment by reproducing white noise and compare these to the simulated magnitude responses obtained above. These are shown in \figref{fig:freq:real}, where the red line is the measured one and the green line is the ideal one (obtained by filtering a discrete-time impulse sequence). The observed deviation is at most 2\,dB, but is inherent to the use of white noise as the input signal. Indeed, by computing the magnitude response of the room using white noise in the simulated computer environment, we obtain similarly random deviations from the flat band (blue line). Therefore we can conclude that the IIR filters designed in the simulated environment are suitable for the real scenario.

\section{Conclusions}\label{sec:conclusion}

In this paper, we described an IIR parametric equalizer design for automatic room equalization at multiple listening points, exploiting a Deep Optimization method based on a spectral loss function and a regularization term which avoids energy unbalance between the speakers. The error is backpropagated to update the neural network weights. We also introduced BiasNet, a novel neural network architecture with learnable input, specifically designed for Deep Optimization. Its behavior is compared to other neural network architectures. We compared BiasNet with two other baseline techniques, one based on FIR filters for frequency deconvolution and another based on a iterative approach for IIR filter design. Compared to the baseline techniques, our method achieves better performance at a remarkably lower computational cost. We also tested the designed IIR filters in a real scenario, showing almost no difference to the simulated one.

The proposed method works offline and assumes a linear environment, therefore other factors, such as the movement of the listening positions are not considered. These require further studies. In the future we plan to perform subjective tests to assess the validity of the approach and devise other metrics based on psychoacoustics, for the optimization. 





\ifCLASSOPTIONcaptionsoff

\fi
\bibliographystyle{IEEEtran}
\bibliography{IEEEabrv,reference}

\begin{thebibliography}{10}
\providecommand{\url}[1]{#1}
\csname url@samestyle\endcsname
\providecommand{\newblock}{\relax}
\providecommand{\bibinfo}[2]{#2}
\providecommand{\BIBentrySTDinterwordspacing}{\spaceskip=0pt\relax}
\providecommand{\BIBentryALTinterwordstretchfactor}{4}
\providecommand{\BIBentryALTinterwordspacing}{\spaceskip=\fontdimen2\font plus
\BIBentryALTinterwordstretchfactor\fontdimen3\font minus
  \fontdimen4\font\relax}
\providecommand{\BIBforeignlanguage}[2]{{%
\expandafter\ifx\csname l@#1\endcsname\relax
\typeout{** WARNING: IEEEtran.bst: No hyphenation pattern has been}%
\typeout{** loaded for the language `#1'. Using the pattern for}%
\typeout{** the default language instead.}%
\else
\language=\csname l@#1\endcsname
\fi
#2}}
\providecommand{\BIBdecl}{\relax}
\BIBdecl

\bibitem{vesa2016equalization}
V.~V\"{a}lim\"{a}ki and J.~D. Reiss, ``All about audio equalization: Solutions
  and frontiers,'' \emph{Applied Sciences}, vol.~6, no.~5, 2016.

\bibitem{reiss2011design}
J.~D. Reiss, ``Design of audio parametric equalizer filters directly in the
  digital domain,'' \emph{Trans. Audio, Speech and Lang. Proc.}, vol.~19,
  no.~6, p. 1843–1848, august 2011.

\bibitem{Karjalainen2005}
M.~{Karjalainen}, T.~{Paatero}, J.~N. {Mourjopoulos}, and P.~D.
  {Hatziantoniou}, ``About room response equalization and dereverberation,'' in
  \emph{IEEE Workshop on Applications of Signal Processing to Audio and
  Acoustics, 2005.}, 2005, pp. 183--186.

\bibitem{cecchi2018room}
S.~Cecchi, A.~Carini, and S.~Spors, ``Room response equalization - a review,''
  \emph{Applied Sciences}, vol.~8, no.~1, 2018.

\bibitem{environments5040044}
Y.~Soeta and Y.~Sakamoto, ``An exploratory analysis of sound field
  characteristics using the impulse response in a car cabin,''
  \emph{Environments}, vol.~5, no.~4, 2018.

\bibitem{HUANG201998}
H.~B. Huang, J.~H. Wu, X.~R. Huang, M.~L. Yang, and W.~P. Ding, ``The
  development of a deep neural network and its application to evaluating the
  interior sound quality of pure electric vehicles,'' \emph{Mechanical Systems
  and Signal Processing}, vol. 120, pp. 98 -- 116, 2019.

\bibitem{cecchi2009automotive}
S.~Cecchi, L.~Palestini, P.~Peretti, F.~Piazza, F.~Bettarelli, and R.~Toppi,
  ``Automotive audio equalization,'' \emph{Journal of the Audio Engineering
  Society}, june 2009.

\bibitem{chang1994inverse}
P.~Chang, C.~G. Lin, and B.~Yeh, ``Inverse filtering of a loudspeaker and room
  acoustics using time-delay neural networks,'' \emph{The Journal of the
  Acoustical Society of America}, vol.~95, no.~6, pp. 3400--3408, 1994.

\bibitem{martinez2018equalization}
M.~A. Martinez~Ramirez and J.~D. Reiss, ``End-to-end equalization with
  convolutional neural networks,'' in \emph{Proceedings of the 21st
  International Conference on Digital Audio Effects (DAFx-18), Aveiro,
  Portugal}, 2018.

\bibitem{AGRAWAL2021107669}
N.~Agrawal, A.~Kumar, V.~Bajaj, and G.~Singh, ``Design of digital {IIR} filter:
  A research survey,'' \emph{Applied Acoustics}, vol. 172, p. 107669, 2021.

\bibitem{krusienski2003pso}
D.~J. {Krusienski} and W.~K. {Jenkins}, ``Adaptive filtering via particle swarm
  optimization,'' in \emph{The Thrity-Seventh Asilomar Conference on Signals,
  Systems Computers, 2003}, vol.~1, 2003, pp. 571--575 Vol.1.

\bibitem{saha2014gsa}
S.~K. Saha, R.~Kar, D.~Mandal, and S.~Ghoshal, ``Gravitation search algorithm:
  Application to the optimal iir filter design,'' \emph{Journal of King Saud
  University - Engineering Sciences}, vol.~26, no.~1, pp. 69 -- 81, 2014.

\bibitem{dodds2020a}
P.~Dodds, ``{A} flexible numerical optimization approach to the design of
  biquad filter cascades,'' \emph{Journal of the Audio Engineering Society},
  october 2020.

\bibitem{jiang2010}
A.~{Jiang} and H.~K. {Kwan}, ``Minimax design of iir digital filters using
  iterative socp,'' \emph{IEEE Transactions on Circuits and Systems I: Regular
  Papers}, vol.~57, no.~6, pp. 1326--1337, 2010.

\bibitem{Sloss2020}
A.~N. Sloss and S.~Gustafson, \emph{2019 Evolutionary Algorithms Review}.\hskip
  1em plus 0.5em minus 0.4em\relax Cham: Springer International Publishing,
  2020, pp. 307--344.

\bibitem{nongpiur2013}
R.~C. {Nongpiur}, D.~J. {Shpak}, and A.~{Antoniou}, ``Improved design method
  for nearly linear-phase iir filters using constrained optimization,''
  \emph{IEEE Transactions on Signal Processing}, vol.~61, no.~4, pp. 895--906,
  2013.

\bibitem{Engel2020DDSP}
J.~Engel, L.~H. Hantrakul, C.~Gu, and A.~Roberts, ``Ddsp: Differentiable
  digital signal processing,'' in \emph{International Conference on Learning
  Representations}, 2020.

\bibitem{kuznetsovdifferentiable}
B.~Kuznetsov, J.~D. Parker, and F.~Esqueda, ``Differentiable {IIR} filters for
  machine learning applications.''

\bibitem{Bhattacharya2020OPTIMIZATIONOC}
P.~Bhattacharya, P.~Nowak, and U.~Z{\"o}lzer, ``Optimization of cascaded
  parametric peak and shelving filters with backpropagation algorithm,'' in
  \emph{Digital Audio Effects (DAFx) 2020, Vienna, Austria}, september 2020.

\bibitem{Nercessian2020Neural}
S.~Nercessian, ``Neural parametric equalizer matching using differentiable
  biquads,'' in \emph{Digital Audio Effects (DAFx) 2020, Vienna, Austria},
  september 2020.

\bibitem{back1991synapses}
A.~D. Back and A.~C. Tsoi, ``{FIR} and {IIR} synapses, a new neural network
  architecture for time series modeling,'' \emph{Neural Computation}, vol.~3,
  no.~3, pp. 375--385, 1991.

\bibitem{Allakhverdiyeva2016nnfilter}
N.~Allakhverdiyeva, ``Application of neural network for digital recursive
  filter design,'' in \emph{2016 IEEE 10th International Conference on
  Application of Information and Communication Technologies (AICT)}, october
  2016, pp. 1--4.

\bibitem{valimaki2019}
V.~{V\"{a}lim\"{a}ki} and J.~{R\"{a}m\"{o}}, ``Neurally controlled graphic
  equalizer,'' \emph{IEEE/ACM Transactions on Audio, Speech, and Language
  Processing}, pp. 1--1, 2019.

\bibitem{ramo2019neural}
J.~R{\"a}m{\"o} and V.~V{\"a}lim{\"a}ki, ``Neural third-octave graphic
  equalizer,'' in \emph{Proceedings of the 22nd International Conference on
  Digital Audio Effects (DAFx-19), Birmingham, UK}, september 2019, pp. 1--6.

\bibitem{PEPE2020107204}
G.~Pepe, L.~Gabrielli, S.~Squartini, and L.~Cattani, ``Evolutionary tuning of
  filters coefficients for binaural audio equalization,'' \emph{Applied
  Acoustics}, vol. 163, p. 107204, 2020.

\bibitem{pepe2020designing}
------, ``Designing audio equalization filters by deep neural networks,''
  \emph{Applied Sciences}, vol.~10, no.~7, p. 2483, 2020.

\bibitem{kirkeby1998freq}
O.~{Kirkeby}, P.~A. {Nelson}, H.~{Hamada}, and F.~{Orduna-Bustamante}, ``Fast
  deconvolution of multichannel systems using regularization,'' \emph{IEEE
  Transactions on Speech and Audio Processing}, vol.~6, no.~2, pp. 189--194,
  march 1998.

\bibitem{ramos2006filter}
G.~Ramos and J.~J. López, ``Filter design method for loudspeaker equalization
  based on {iir} parametric filters,'' \emph{Journal of the Audio Engineering
  Society}, vol.~54, no.~12, pp. 1162--1178, december 2006.

\bibitem{behrends2011automatic}
H.~Behrends, A.~Von~dem Knesebeck, W.~Bradinal, P.~Neumann, and U.~Zölzer,
  ``Automatic equalization using parametric {IIR} filters,'' \emph{Journal of
  the Audio Engineering Society}, vol.~59, no.~3, pp. 102--109, march 2011.

\bibitem{pepegravitational}
G.~Pepe, L.~Gabrielli, S.~Squartini, L.~Cattani, and C.~Tripodi,
  ``Gravitational search algorithm for iir filter-based audio equalization,''
  in \emph{2020 28th European Signal Processing Conference (EUSIPCO),
  Amsterdam}, 2020.

\bibitem{zolzer2011dafx}
U.~Z{\"o}lzer, \emph{DAFX: digital audio effects}.\hskip 1em plus 0.5em minus
  0.4em\relax John Wiley \& Sons, 2011.

\bibitem{lopez2018easing}
D.~Lopez-Paz and L.~Sagun, ``Easing non-convex optimization with neural
  networks,'' in \emph{International Conference on Learning Representations
  (ICLR 2018)}, 2018.

\bibitem{Oppenheim}
A.~V. Oppenheim and R.~V. Schafer, \emph{Discrete-time Signal
  Processing}.\hskip 1em plus 0.5em minus 0.4em\relax Upper Saddle River, NJ,
  USA: Prentice-Hall, Inc., 1999.

\bibitem{caracalla2017gradient}
H.~Caracalla and A.~Roebel, ``Gradient conversion between time and frequency
  domains using wirtinger calculus,'' in \emph{Digital Audio Effects (DAFx)
  2017, Edinburgh, United Kingdom}, september 2017.

\bibitem{Hooke1961}
R.~Hooke and T.~A. Jeeves, ``{"Direct Search"} solution of numerical and
  statistical problems,'' \emph{J. ACM}, vol.~8, no.~2, p. 212–229, april
  1961.

\bibitem{LEWIS2000191}
R.~M. Lewis, V.~Torczon, and M.~W. Trosset, ``Direct search methods: then and
  now,'' \emph{Journal of Computational and Applied Mathematics}, vol. 124,
  no.~1, pp. 191 -- 207, 2000.

\bibitem{farina2007advancements}
A.~Farina, ``Advancements in impulse response measurements by sine sweeps,''
  \emph{Journal of the Audio Engineering Society}, may 2007.

\bibitem{sitzmann2020implicit}
V.~Sitzmann, J.~Martel, A.~Bergman, D.~Lindell, and G.~Wetzstein, ``Implicit
  neural representations with periodic activation functions,'' vol.~33, 2020.

\bibitem{kingma2014method}
D.~P. Kingma and J.~Ba, ``Adam: a method for stochastic optimization,'' in
  \emph{3rd International Conference for Learning Representations, San Diego,
  2015}, 2015.

\end{thebibliography}

\end{document}